\documentclass[journal=macromol,manuscript=article,layout=twocolumn]{achemso}

\usepackage[version=3]{mhchem} % Formula subscripts using \ce{}

\author{Gauthier Legrand}
\affiliation{ENSL, CNRS, Laboratoire de physique, F-69342 Lyon, France}
\author{Sébastien Manneville}%
\affiliation{ENSL, CNRS, Laboratoire de physique, F-69342 Lyon, France}
\alsoaffiliation{Institut Universitaire de France (IUF)}
\author{Gareth H. McKinley}
\affiliation{Hatsopoulos Microfluids Laboratory, Department of Mechanical Engineering, MIT, 77 Massachusetts Avenue,
Cambridge, Massachusetts 02139, USA}
\author{Thibaut Divoux}%
\affiliation{ENSL, CNRS, Laboratoire de physique, F-69342 Lyon, France}
\email{Thibaut.Divoux@ens-lyon.fr}

\title[An \textsf{achemso} demo]
{Dual origin of viscoelasticity in polymer-carbon black hydrogels: a rheometry and electrical spectroscopy study}
\abbreviations{IR,NMR,UV}

\usepackage{siunitx} %%%%% TO BE REMOVED ??? %%%%% 
\usepackage[dvipsnames]{xcolor}
\usepackage{amssymb} 
\definecolor{blue6}{RGB}{83, 177, 223}
\definecolor{blue8}{RGB}{65, 137, 207}
\definecolor{blue10}{RGB}{46, 98, 190}

\definecolor{cmc2}{RGB}{0, 0, 0}
\definecolor{cmc3}{RGB}{159, 100, 63}
\definecolor{cmc4}{RGB}{212, 133, 85}
\definecolor{cmc5}{RGB}{255, 166, 106}

\begin{document}

%%%%%%%%%%%%%%%%%%%%%%%%%%%%%%%%%%%%%%%%%%%%%%%%%%%%%%%%%%%%%%%%%%%%%
%% The "tocentry" environment can be used to create an entry for the
%% graphical table of contents. It is given here as some journals
%% require that it is printed as part of the abstract page. It will
%% be automatically moved as appropriate.
%%%%%%%%%%%%%%%%%%%%%%%%%%%%%%%%%%%%%%%%%%%%%%%%%%%%%%%%%%%%%%%%%%%%%
%\begin{tocentry}

%Some journals require a graphical entry for the Table of Contents.
%This should be laid out ``print ready'' so that the sizing of the text is correct.

%Inside the \texttt{tocentry} environment, the font used is Helvetica 8\,pt, as required by \emph{Journal of the American Chemical
%Society}.

%The surrounding frame is 9\,cm by 3.5\,cm, which is the maximum permitted for  \emph{Journal of the American Chemical Society}
%graphical table of content entries. The box will not resize if the content is too big: instead it will overflow the edge of the box.

%This box and the associated title will always be printed on a separate page at the end of the document.

%\end{tocentry}

%%%%%%%%%%%%%%%%%%%%%%%%%%%%%%%%%%%%%%%%%%%%%%%%%%%%%%%%%%%%%%%%%%%%%
%% The abstract environment will automatically gobble the contents
%% if an abstract is not used by the target journal.
%%%%%%%%%%%%%%%%%%%%%%%%%%%%%%%%%%%%%%%%%%%%%%%%%%%%%%%%%%%%%%%%%%%%%
\begin{abstract}
 Nano-composites formed by mixing nanoparticles and polymers offer a limitless creative space for the design of functional advanced materials with a broad range of applications in materials and biological sciences. Here we focus on aqueous dispersions of hydrophobic colloidal soot particles, namely carbon black (CB) dispersed with a sodium salt of carboxymethylcellulose (CMC), a food additive known as cellulose gum that bears hydrophobic groups, which are liable to bind physically to CB particles. Varying the relative content of CB nanoparticles and cellulose gum allows us to explore a rich phase diagram that includes a gel phase observed for large enough CB content. We investigate this hydrogel using rheometry and electrochemical impedance spectroscopy. CB-CMC hydrogels display two radically different types of mechanical behaviors that are separated by a critical CMC-to-CB mass ratio $r_c$. For $r<r_c$, i.e., for low CMC concentration, the gel is electrically conductive and shows a glassy-like viscoelastic spectrum, pointing to a microstructure composed of a percolated network of CB nanoparticles decorated by CMC. In contrast, gels with CMC concentration larger than $r_c$ are non-conductive, indicating that the CB nanoparticles are dispersed in the cellulose gum matrix as isolated clusters, and act as physical crosslinkers of the CMC network, hence providing mechanical rigidity to the composite. Moreover, in the concentration range, $r>r_c$ CB-CMC gels display a power-law viscoelastic spectrum that  depends strongly on the CMC concentration. These relaxation spectra can be rescaled onto a master curve that exhibits a power-law scaling in the high-frequency limit, with an exponent that follows Zimm theory, showing that CMC plays a key role in the gel viscoelastic properties for $r>r_c$.
 Our results offer an extensive experimental characterization of CB-CMC dispersions that will be useful for designing soft nano-composites based on hydrophobic interactions. 
\end{abstract}

%%%%%%%%%%%%%%%%%%%%%%%%%%%%%%%%%%%%%%%%%%%%%%%%%%%%%%%%%%%%%%%%%%%%%
%% Start the main part of the manuscript here.
%%%%%%%%%%%%%%%%%%%%%%%%%%%%%%%%%%%%%%%%%%%%%%%%%%%%%%%%%%%%%%%%%%%%%
\section{Introduction}

Carbon black (CB) particles are colloidal soot particles produced from the incomplete combustion of fossil fuels. These textured nanoparticles of typical size 0.5~$\mu$m are made of permanently fused ``primary” particles of diameter 20-40~nm\cite{Samson:1987}. Cheap and industrially produced at large scale, they are broadly employed for their mechanical strength, high surface area, and their electrical conductive properties, even at low volume fractions. Applications include pigments for ink \cite{Spinelli:1998}, reinforcing fillers in tires and other rubber products \cite{EhrburgerDolle:2001,Huang:2002}, electrically conductive admixture in cement \cite{Wen:2007}, conductive materials for supercapacitors \cite{Pandolfo:2006}, biosensors \cite{Silva:2017,Kour:2020}, and electrodes for semi-solid flow batteries \cite{Duduta:2011,Li:2013b,Youssry:2013,Narayanan:2017,Narayanan:2021}. Being much cheaper than carbon nanotubes or graphene, CB nanoparticles appear promising for applications in energy storage \cite{Khodabakhshi:2020}, including flow electrodes for which the ultimate goal is to maximize the conductivity, while minimizing the shear viscosity of the material. In that framework, flow batteries based on aqueous dispersion of carbon black nanoparticles have recently received an upsurge of interest \cite{Li:2013b,Chou:2014,Parant:2017,Youssry:2018,Meslam:2022}.  

Due to their hydrophobic properties, carbon black particles are easily dispersed in aprotic solvents such as hydrocarbons \cite{Waarden:1950}, where particles interact only via van der Waals forces \cite{Hartley:1985} that correspond to a short-range attractive potential, whose depth is typically about 30~$k_B$T in light mineral oil~\cite{Trappe:2007}. As a result, carbon black dispersions organize into space-spanning networks even at low volume fractions, and behave as soft gels \cite{Trappe:2000,Trappe:2001,Kawaguchi:2001} characterized by a yield stress at rest and a highly time-dependent mechanical response under external shear, which involves delayed yielding, heterogeneous flows \cite{Gibaud:2010,Sprakel:2011,Grenard:2014,Gibaud:2016}, and shear-induced memory effects \cite{Ovarlez:2013,Divoux:2013,Helal:2016}. 

In contrast, untreated CB particles are difficult to disperse in water where they tend to flocculate rapidly, before creaming or sedimenting \cite{Li:2005,Parant:2017}. Stabilizing aqueous dispersions of CB nanoparticles requires keeping the particles apart, either by electrostatic repulsion or by steric hindrance. In practice, this is achieved in three different ways: ($i$) surface oxidation yielding acidic functional groups \cite{Paredes:2005}, ($ii$) functionalization of CB particles with polymers, i.e., polymer grafting chemically onto their surface \cite{Liu:2003,Liu:2006,Lin:1998,Tsubokawa:2002,Yang:2007,Pei:2014,Wang:2019} or CB encapsulation through emulsion polymerization \cite{Tiarks:2001,Casado:2007}, and ($iii$) physical adsorption of a polymer dispersant. 
The latter method allows reaching CB mass fractions in water as large as 20\%, and the dispersants investigated include polyelectrolytes \cite{Alves:1981}, ionic surfactants \cite{Ijima:2013,Hanada:2013} such as sulfonate surfactants \cite{Ogura:1993,Ogura:1994,Gupta:2005,Zhao:2013}, sulfonic acids \cite{Mizukawa:2009,Subramanian:2021}, cetyltrimethylammonium bromide (CTAB) \cite{Bele:1998,Gupta:2005,Porcher:2010,Eisermann:2014} and chloride (CTAC) \cite{Ridaoui:2006}, non-ionic surfactants \cite{Gupta:2005,Sis:2009} such as silicone surfactants \cite{Lin:2002} or block copolymers surfactants \cite{Miano:1992,Sis:2009,Yasin:2012}, as well as biopolymers such as Arabic gum \cite{Parant:2017,Ngouamba:2020}, or polysaccharides \cite{Guerfi:2007,Lee:2008,Porcher:2009,Garcia:2018}. From a structural point of view, dispersants adsorb as monolayers onto the surface of CB nanoparticles due to hydrophobic interactions, whose strength depends on the molecular structure and weight of the dispersant  \cite{Porcher:2009,Zhao:2013}. Irrespective of the nature of the dispersant, such stabilized CB dispersions behave as shear-thinning fluids \cite{Barrie:2004,Lee:2008,Garcia:2018}. However, depending on the formulation, carbon black polymer mixtures may present a solid-like behavior at rest with weakly time-dependent properties \cite{Porcher:2009,Yasin:2012,Subramanian:2021}, suggesting the presence of a percolated network of the CB nanoparticles \cite{Aoki:2003}.

The large variety of dispersants used so far in CB dispersions is in stark contrast with the limited knowledge regarding the link between the microstructure and rheological properties of the resulting materials. Among the open questions, it remains to disentangle the respective contributions of the CB and the dispersant to the macroscopic mechanical properties of the mixture, and whether the CB nanoparticles form a percolated network on their own, or if they are bridged by the polymeric chains.  

Here we perform such an in-depth study with a semi-flexible anionic polysaccharide dispersant, namely carboxymethylcellulose (CMC). CMC is a water soluble cellulose ether, which is commonly used as a water binder and thickener in pharmacy, cosmetics, food products \cite{Rahman:2021}, and as a dispersing agent in semi-solid flow batteries \cite{Lee:2008}, while showing a great potential for biomedical applications \cite{Jiang:2009}. The solubility and overall properties of CMC are set primarily by its molecular weight, and to a lesser extent by its degree of substitution (DS) \cite{Kulicke:1996,Clasen:2001}. The latter is defined as the number of hydrogen atoms in hydroxyl groups of glucose units replaced by carboxymethyl, and varies typically between 0.4 and 3 \cite{Rahman:2021}. Over a broad range of concentrations, CMC aqueous solutions are shear-thinning viscoelastic fluids \cite{Ghannam:1997,Benchabane:2008}. However, weakly substituted CMC, i.e., with DS values lower than about 1, display hydrophobic interactions that favor inter-chain association in aqueous solution, yielding larger viscosities, and eventually leading to a sol-gel transition for large enough polymer concentrations \cite{Elliott:1971,Barba:2002,Medronho:2012,Glasser:2012,Lopez:2018,Lopez:2021}. 

In the present work, we take advantage of such hydrophobic regions on CMC molecules to use this polymer as a dispersant of CB nanoparticles. Varying the contents of CB and CMC, we unravel a rich phase diagram, which we characterize by rheometry and electrochemical impedance spectroscopy. The outline of the manuscript is as follows: after presenting the materials and methods, we introduce the phase diagram and focus specifically on the hydrogel phase. We then discuss the impact of the CMC concentration on the gel elastic properties before turning to the role of the CB content. Our results allow us to identify two different types of hydrogels, whose microstructures are sketched and extensively discussed before concluding. 

\section{Materials and methods}

Samples are prepared by first dissolving sodium carboxymethyl cellulose (Sigma Aldrich, $M_w=250$~kg.mol$^{-1}$ and DS~$=0.9$) in deionized water. Stock solutions up to 5\% wt.~are prepared and stirred at room temperature for 48 hours until homogeneous, before adding the CB nanoparticles (VXC72R, Cabot). Samples are placed in a sonicator bath for two rounds of $90$~min separated by a period of 24~h under mechanical stirring. The samples are finally left at rest for another 24~h before being tested. The CMC solution is considered to be the solvent, while CB particles is the dispersed phase; hence we define the CMC weight concentration as $c_{\rm CMC} = m_{\rm CMC} / (m_{\rm CMC} + m_{\rm water})$ and the CB weight fraction as $x_{\rm CB} = m_{\rm CB} / (m_{\rm CB} + m_{\rm CMC} + m_{\rm water})$, where $m_{\rm CB}$, $m_{\rm CMC}$, and $m_{\rm water}$ are respectively the mass of CB, CMC, and water in the sample. 

Rheological measurements are performed with a stress-controlled rheometer (MCR 302, Anton Paar) equipped with a cone-and-plate geometry (angle {2}$^{\circ}$, radius {20}~mm). For very soft samples with elastic moduli lower than 10~Pa, we use a larger cone-and-plate geometry (angle {2}$^{\circ}$, radius {25}~mm), whereas for the stiffest samples with elastic moduli larger than $10$~kPa, we use a parallel-plate geometry (gap {1}~mm, radius {20}~mm). For all the geometries used, the stator is smooth and the rotor is sandblasted. 
Finally, to ensure a reproducible initial state following the loading step into the shear cell of the rheometer, each sample is shear-rejuvenated at $\dot \gamma=$~{500}~s$^{-1}$ (or $\dot \gamma=$~{50}~s$^{-1}$ for samples with elastic moduli larger than $10$~kPa), before being left at rest for 1200~s, during which we monitor the linear viscoelastic properties through small amplitude oscillatory shear ($\gamma_0=$~{0.03-0.3}\%, and $f=1$~Hz, see supplemental Fig.~\ref{Sfig:time_sweep}).

\begin{figure}[!t]
    \centering
    \includegraphics[width=0.95\linewidth]{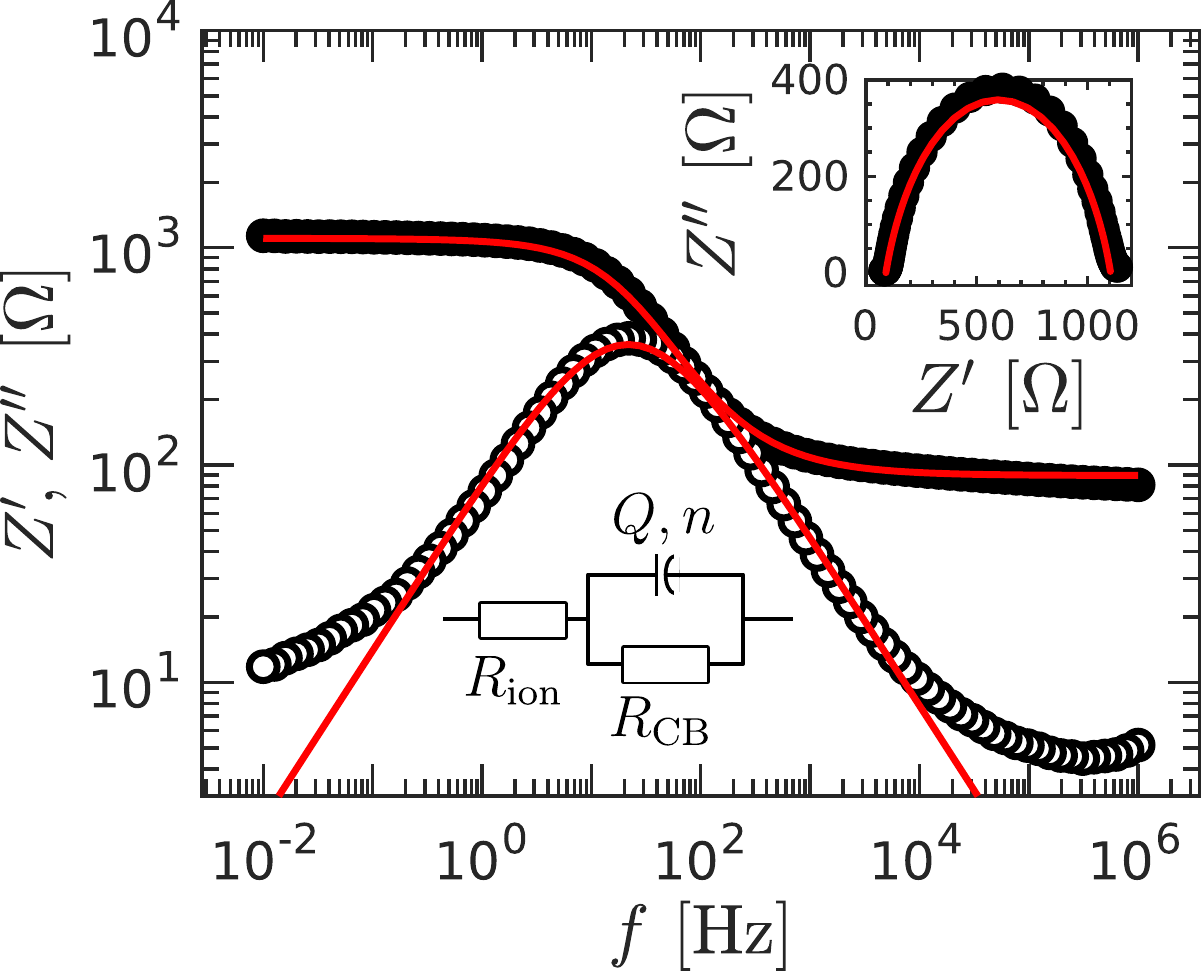}
    \caption{\textbf{Frequency dependence of the resistance $Z'$ ($\bullet$) and reactance $Z''$ ($\circ$) of a CMC-CB dispersion}. Inset: Nyquist plot $Z''$ vs.~$Z'$ for the same data. Measurement performed in AC mode by ramping down the frequency from $10^6$~Hz to $10^{-2}$~Hz. Each point is averaged over two cycles. The red continuous curves in both the main graph and the inset show the best fit of the data for $Z'$ and $Z''$ simultaneously by Eq.~\eqref{eq:elec}, which corresponds to the electrical circuit sketched in the main graph with $R_{\rm ion}=90$~$\Omega$, $R_{\rm CB}=1.0$~k$\Omega$, $n=0.8$, and $Q=2.1\times 10^{-5}$~$\Omega^{-1}$.s$^n$. Data obtained on a sample composed of $c_{\rm CMC}=0.15$\%~wt.~and $x_{\rm CB}=8$\%~wt.}
    \label{fig:bode}
\end{figure}

\begin{figure*}[!t]
    \centering
    \includegraphics[width=0.9\textwidth]{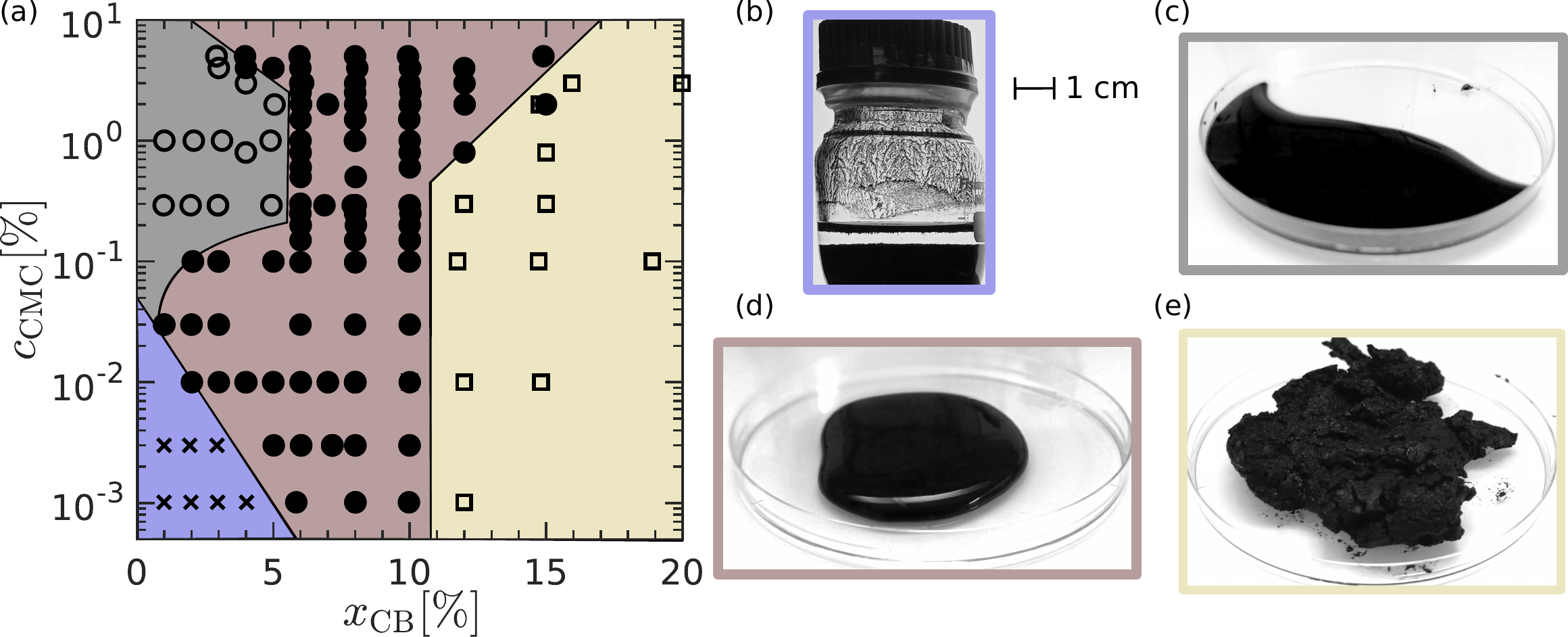}
    \caption{(a) \textbf{Phase diagram of aqueous CMC-CB dispersions} as a function of the CB solid weight fraction $x_{\rm CB}$ and the CMC weight fraction $c_{\rm CMC}$. Pictures of (b)~demixed phase  [blue region and $\times$ symbols in (a)], (c)~the viscoelastic liquid phase [grey region and $\circ$ symbols in (a)], (d)~the viscoelastic solid phase [red region and $\bullet$ symbols in (a)], (e)~brittle paste phase [yellow region and $\square$ symbols in (a)].}
    \label{fig2}
\end{figure*}  

Electrical measurements are performed in AC mode  in a cylindrical cell made of Teflon (thickness of about 1~mm, and surface of about 0.5~cm$^2$). The inner sides are made of metal act as electrodes that are connected to a multi-frequency impedance analyzer (SP-300 Potentiostat, Biologic). A decreasing ramp of frequency allows determining the frequency dependence of the sample impedance $Z^*(\omega)= Z' - iZ''$, which real and imaginary part, $Z'$ and $Z''$ respectively, are shown in Fig.~\ref{fig:bode} for a sample containing $c_{\rm CMC}=0.15$\%~wt.~and $x_{\rm CB}=8$\% wt. The resistance $Z'$ shows a decreasing step shape, while the reactance $Z''$ displays a bell-shaped curve. Overall, the complex impedance measured experimentally can be well fitted by the following equation:
\begin{equation}
    Z^*(\omega)=R_{\rm ion} + \frac{R_{\rm CB}}{1+R_{\rm CB}Q(i\omega)^n}
\label{eq:elec}
\end{equation}
that corresponds to the simple circuit sketched in Fig.~\ref{fig:bode}. Equation~\eqref{eq:elec} accounts for the additive contributions of the ions in solution, and an electronically-conductive percolated network of CB particles. 
More precisely, the circuit comprises a resistance $R_{\rm ion}$ accounting for the ionic conductivity of the sample in series with two elements modeling the percolated network of CB particles, namely a resistance $R_{\rm CB}$ in parallel with a constant phase element characterized by a dimensionless exponent $n$, and a parameter $Q$. The latter element, which was first introduced in the context of particulate suspensions \cite{Cole:1928} relates to fractal interfaces and corresponds to an imperfect capacitor \cite{Macdonald:2018,Lasia:2022}. Although such a modeling fails to describe the dependence of $Z''(\omega)$ at very low and very high frequencies, potentially due to some parasitic inductance, it does accounts very well for $Z'(\omega)$ over eight orders of magnitude. The low- and high-frequency limits of $Z'(\omega)$ allow us to determine the resistances $R_{\rm ion}$ and $R_{\rm CB}$. The latter parameter can be converted into the electrical conductivity $\sigma_{\rm CB}=k/R_{\rm CB}$, with $k$ the cell constant determined by independent measurements on KCl solutions of different concentrations (here $k=0.38\pm0.01$~cm$^{-1}$, see supplemental Fig.~\ref{Sfig:calibration}).

\section{Results and discussion}

\subsection{Phase diagram}

We first discuss qualitatively the outcome of dispersing carbon black (CB) nanoparticles in a solution of carboxymethylcellulose (CMC). In practice, we observe four different phases that are summarized in the phase diagram reported in Fig.~\ref{fig2}. For low CB and CMC concentration, the aqueous dispersion is unstable and the CB nanoparticles sediment within about 20~min [Fig.~\ref{fig2}(b)]. On the one hand, increasing the CMC concentration beyond about $10^{-2}$\% allows stabilizing the CB nanoparticles, yielding a viscoelastic liquid [Fig.~\ref{fig2}(c)]. On the other hand, increasing the content in CB nanoparticles confers gel-like properties upon the samples, i.e., the sample shows a solid-like behavior at rest and for small deformations, while it flows for large enough stresses [Fig.~\ref{fig2}(d)]. Finally, for CB content larger than about 10\% wt., the sample behaves as a strongly elastic paste, with a fragile behavior to the touch [Fig.~\ref{fig2}(e)].

\begin{figure}[!t]
    \centering
    \includegraphics[width=0.95\columnwidth]{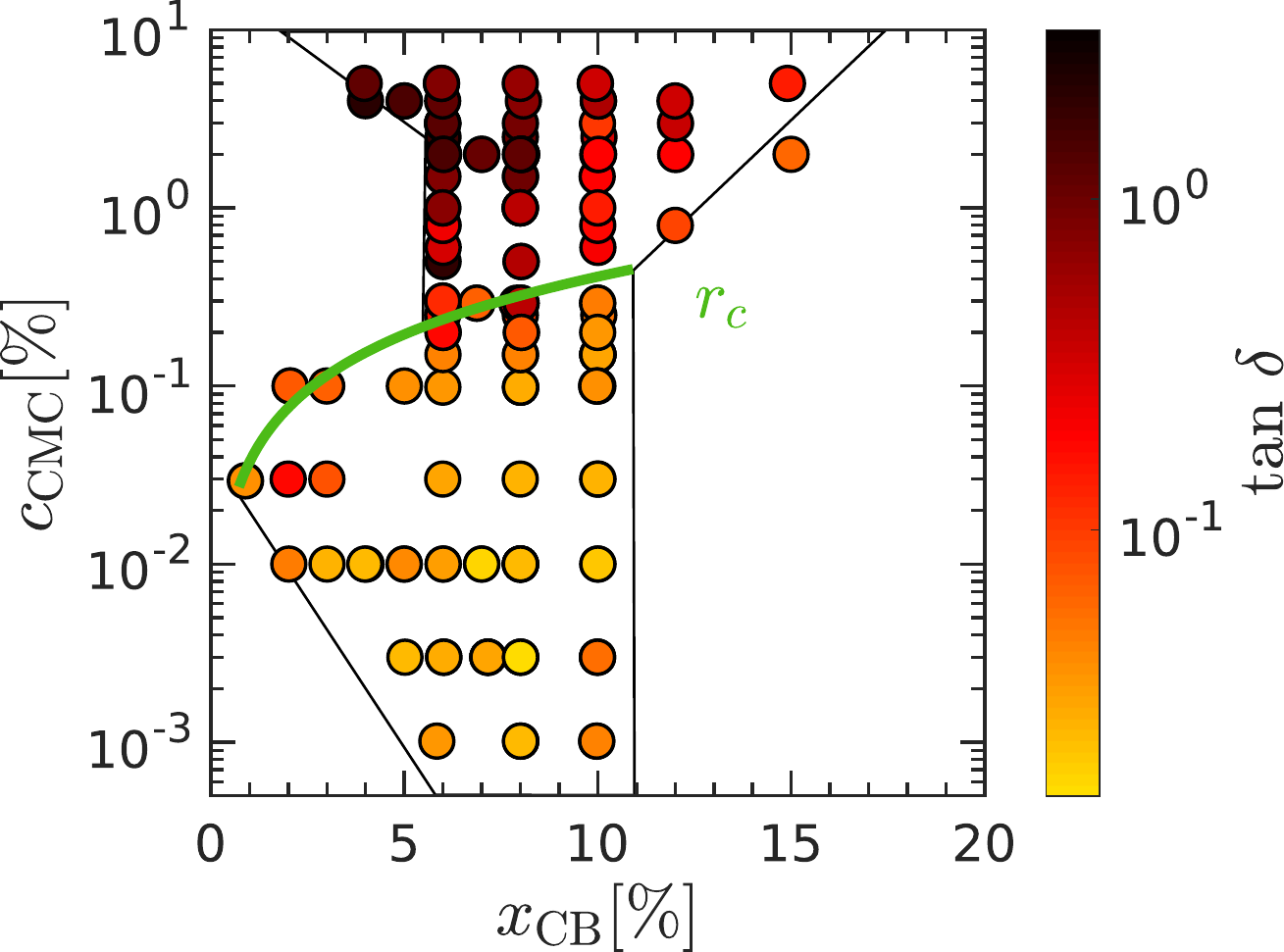}
    \caption{\textbf{Hydrogel region of the phase diagram of aqueous CMC-CB dispersions} as a function of the CB solid weight fraction $x_{\rm CB}$ and the CMC weight fraction $c_{\rm CMC}$. Color levels code for the loss factor $\tan \delta=G''/G'$ of the gel phase determined by small amplitude oscillatory shear at $f=1$~Hz. The green curve corresponds to $r=r_c$, and separates two regions in the gel phase with samples of different microstructures.}
    \label{fig3}
\end{figure}

In the present work, we focus on the hydrogel phase, which is observed over the entire range of CMC concentrations explored, and for CB content ranging between a few \% wt.~and about 12\% wt. In order to quantify the gel rheological properties, we measure its linear viscoelastic properties through small amplitude oscillatory shear, as detailed above. The ratio of the viscous to the elastic modulus measured at $f=1$~Hz, i.e., $G''/G'=\tan \delta$, also known as the loss factor, is reported in Fig.~\ref{fig3} (see supplemental Figs.~\ref{Sfig:diagramme_gseconde} and \ref{Sfig:diagramme_tandelta} for a similar representation of $G'$ and $G''$, respectively). Such a phase diagram built upon the loss factor highlights two different regions, which correspond to samples that mainly differ by their CMC concentration. Samples with a lower CMC concentration display relatively less viscous dissipation ($\tan \delta \lesssim 0.1$) than samples with the highest CMC concentration ($\tan \delta \gtrsim 0.1$). This observation suggests that CMC-CB hydrogels come in two different flavors, depending on the polymer content. We quantify these qualitative results by studying the scaling of the viscoelastic and electrical properties of the samples, with respect to the CMC concentration and the CB content.

\begin{figure}[t!]
    \centering
    \includegraphics[width=0.95\columnwidth]{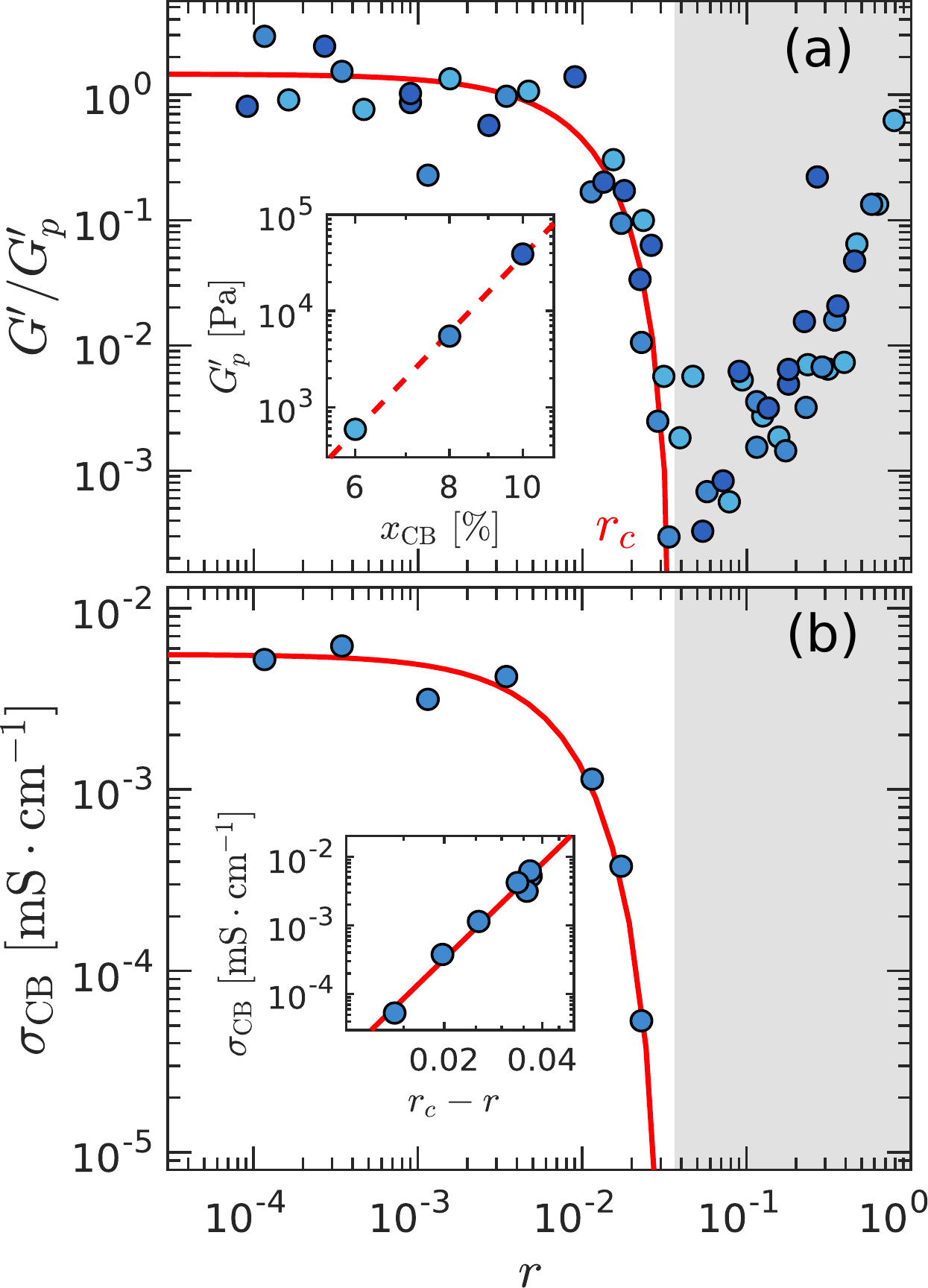}
    \caption{\textbf{Evidence for a transition between two rheological and electrical regimes.} (a) Normalized elastic modulus $G'/G'_p$ measured at $f = 1$~Hz vs.~mass ratio $r=m_{\rm CMC}/m_{\rm CB}$. Each data point has been measured 1200~s after a shear rejuvenation step (see section ``materials and methods"). Color encodes the CB content $x_{\rm CB}$: 6\% (\textcolor{blue6}{$\bullet$}), 8\% (\textcolor{blue8}{$\bullet$}), and 10\% (\textcolor{blue10}{$\bullet$}). The plateau elastic modulus $G'_p$ is defined as the average value of $G'$ over the range $10^{-4}\leqslant r\leqslant 10^{-2}$. The continuous red curve is the best power-law fit of $G'/G'_p$ vs.~$|r-r_c|$ with $r_c=0.037$, yielding an exponent $3.8\pm 0.5$. Inset: $G'_p$ vs.~$x_{CB}$. The red dashed line is the best power-law fit of the data, yielding an exponent $8.2\pm0.6$. Inset: $G'_p$ vs $x_{\rm CB}$. (b) Electrical conductivity $\sigma_{\rm CB}$ of the CMC-CB dispersions vs.~mass ratio $r=m_{\rm CMC}/m_{\rm CB}$, with $x_{\rm CB}=8$\%~wt. The continuous red curve is the best power-law fit of $\sigma_{\rm CB}$ vs.~$|r-r_c|$ with $r_c=0.037$, yielding an exponent $4.6\pm 0.5$. }
    \label{fig4}
\end{figure}

\subsection{Impact of CMC concentration on the gel elastic properties}

In this section, we first discuss the impact of the CMC concentration at fixed CB content, which corresponds to a vertical cut in the phase diagram reported in Fig.~\ref{fig2}(a). The dependence of $G'$ with the CMC concentration is pictured in Fig.~\ref{fig4}(a), as a function of the mass ratio $r=m_{\rm CMC}/m_{\rm CB} = c_{\rm CMC} \left( 1-x_{\rm CB} \right) / x_{\rm CB}$, which represents the effective number of CMC molecules per CB nanoparticles. At low CMC concentrations, we observe that the elastic modulus $G'$ is constant $G'=G'_p$, independent of $r$ over about two decades of CMC concentration, i.e., $10^{-4}\leqslant r\leqslant 10^{-2}$. For $r>10^{-2}$, the elastic modulus drops abruptly by about 3 orders of magnitude for increasing CMC concentration within a narrow range of $r$ values, reaching a minimum value at $r=r_c \simeq 0.037$. Finally, for $r>r_c$, increasing the CMC concentration translates into an increase of $G'$, which scales roughly as a power-law function of $r$. This evolution of $G'$ over the whole range of $r$ is robust, as evidenced by the data reported in Fig.~\ref{fig4}(a) for three different CB content, namely $x_{CB}=6$, 8 and 10\%, as emphasized in Fig.~\ref{Sfig:Gprime_vs_cmc_c} in the supplemental material.

\begin{figure}[!t]
    \centering
    \includegraphics[width=0.9\columnwidth]{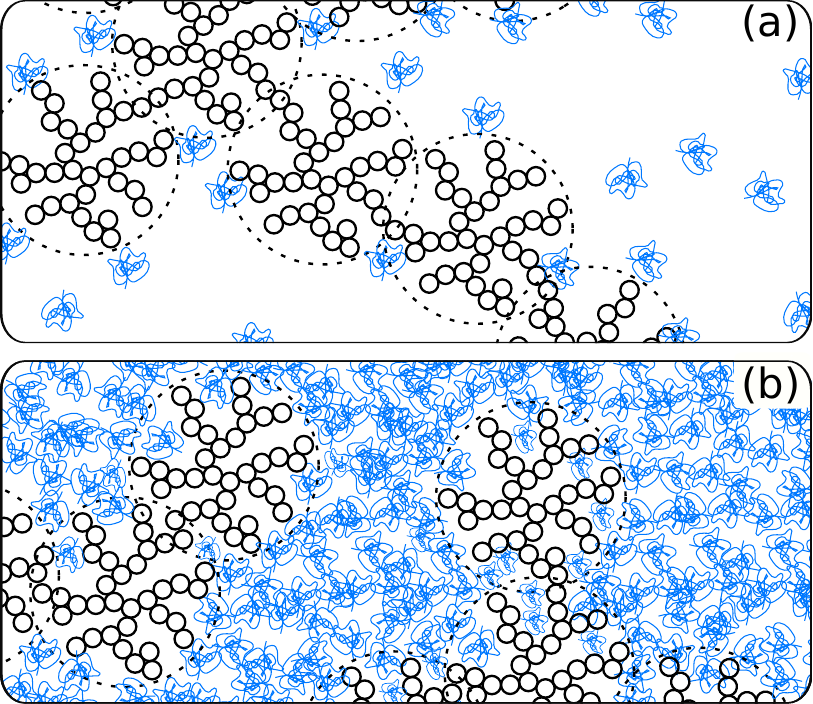}
    \caption{\textbf{Sketch of CMC-CB hydrogels microstructure} in (a) the colloid-dominated regime $r < r_c$, and (b) the polymer-dominated regime $r>r_c$. The black solid circles represent the primary CB nodules, which form the unbreakable fractal CB particles that are individually identified by dashed circles, while the blue coils represent CMC macromolecules.}
    \label{fig:sketch_microstructure}
\end{figure}

These observations unambiguously confirm the trends determined thanks to the loss factor and show that the linear elastic properties of CMC-CB hydrogels have two distinct origins depending on the relative content in CB and CMC. For $r<r_c$, the gel elastic properties are set by the amount of CB nanoparticles [see inset in Fig.~\ref{fig4}(a)] and independent of the CMC concentration, whereas for $r>r_c$, the elastic modulus is an increasing function of the CMC concentration, irrespective of the CB content. Moreover, these results suggest that the gel microstructure is drastically different on each side of $r_c$. To get more insights on the hydrogel microstructure, we build upon the fact that the CB nanoparticles are electrically conductive. 
\begin{figure*}[!t]
    \centering
    \includegraphics[width=0.95\textwidth]{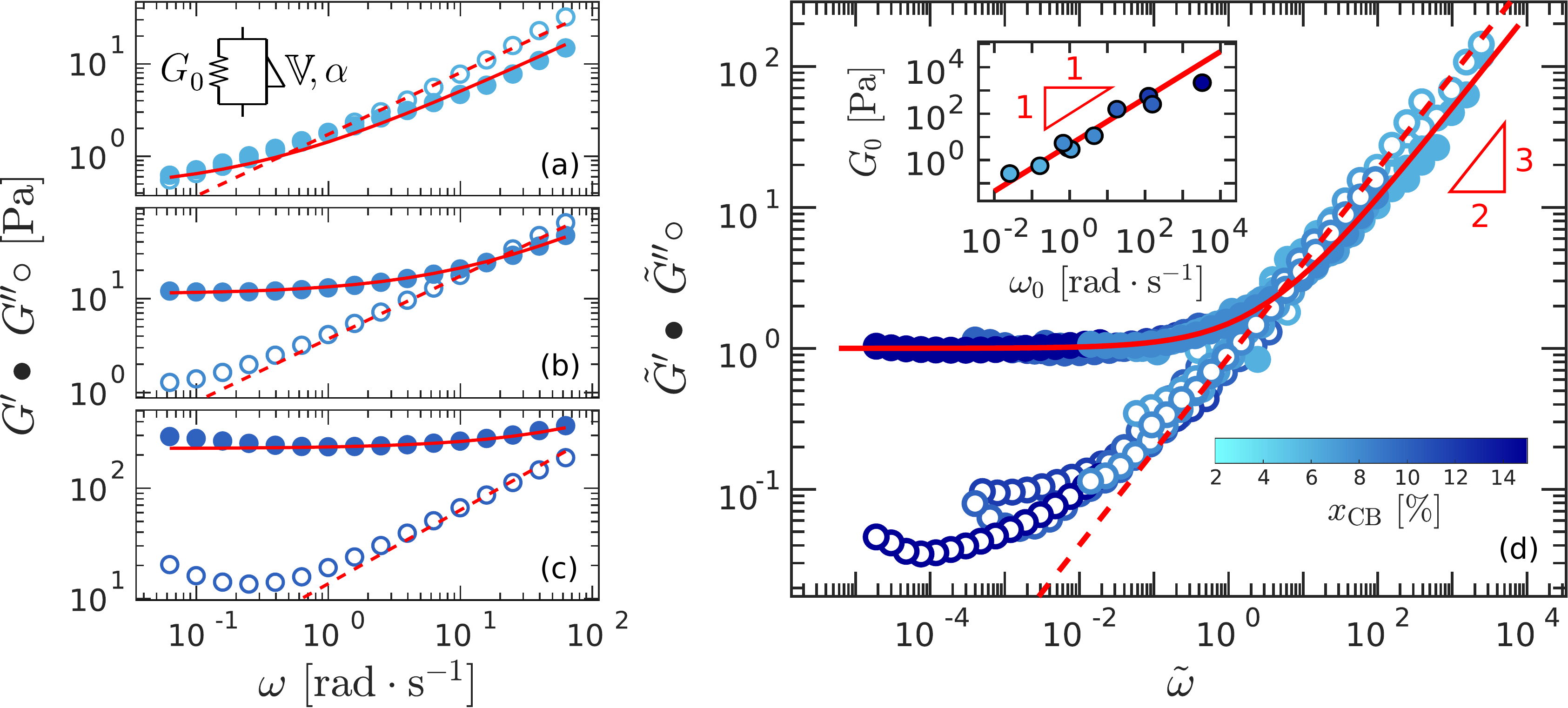}
    \caption{\textbf{Viscoelastic spectrum in the polymer-dominated regime.} Frequency dependence of the elastic and viscous moduli, $G'$ and $G''$ resp., of CMC-CB dispersions with three different CB contents $x_{\rm CB}=$  (a) 6\%, (b) 8\%, and (c) 10\% with $c_{\rm CMC}=2$\%. The red curves are the best fits of the data to a fractional Kelvin-Voigt (FKV) model [see Eq.~\eqref{eq:FKV}], which is sketched in (a). (d) Master curve for the frequency dependence of the viscoelastic moduli of CMC-CB dispersions, obtained by normalizing both the moduli and the frequency: $\tilde G'=G'/G_0$, $\tilde G''=G''/G_0$, and $\tilde \omega = \omega/\omega_0$ with $\omega_0=(G_0/\mathbb{V})^{1/\alpha}$ for different CB contents ranging from 6\% to 15\% with $c_{\rm CMC}=2$\%. The red curves correspond to the normalized FKV model $\tilde{G}^\star = 1 + \left( i \tilde{\omega} \right)^\alpha$ with $\alpha = 2/3$. Inset: $G_0$ vs.~$\omega_0$; the red continuous line is the best linear fit of the data.}
    \label{fig6}
\end{figure*}

We have performed AC electrical measurements on a series of gels with a fixed CB content ($x_{\rm CB}=8$\% wt.) and over a broad range of CMC concentrations. For each gel, we determine its electrical conductivity $\sigma_{\rm CB}$, associated with the possible percolated network of CB particles (see section Material and methods for details about electrical measurements). 
The data are reported in Fig.~\ref{fig4}(b). We observe that the electrical conductivity shows a similar dependence upon the CMC concentration to that of the elastic modulus. For $r<r_c$, the electric conductivity is high, i.e., $\sigma_{\rm e} \simeq 10$~mS.cm$^{-1}$, independent of the CMC concentration, whereas for $r>r_c$, the electric conductivity drops by 3 orders of magnitude down to a negligible value following a power-law of $r_c-r$, with an exponent 4.6 [see inset in Fig.~\ref{fig4}(b)]. 

These results demonstrate that CB nanoparticles form a space-spanning network for $r<r_c$, turning the sample into an electrically conductive material. In contrast, the negligible electrical conductivity observed for $r>r_c$ points towards a microstructure in which the CB nanoparticles are isolated as individual particles or clusters. We can further conclude that these isolated particles or clusters serve as physical cross-linker, which provide sold-like properties to the CMC matrix, for CMC dispersions alone behave as viscoelastic liquids regardless of the CMC concentration \cite{Ghannam:1997,Benchabane:2008}. These two different microstructures are sketched in Fig.~\ref{fig:sketch_microstructure}. 
Finally, note that the transition from a gel network in which the elasticity is set by the CB nanoparticle alone to a gel network in which the elasticity results from the polymer network physically cross-linked by CB nanoparticles occurs over a narrow range of $r$ values. The  value $r_c \simeq 0.037$ corresponds to the critical amount of polymer required to decorate all the CB nanoparticles, such that further addition of CMC only viscosifies the solvent.    

Having identified two different microstructures on each side of $r_c$, we now determine in detail their respective linear viscoelastic properties, namely their linear viscoelastic spectrum and the impact of the CB content.  

\subsection{Polymer-dominated regime ($r>r_{\rm c}$)}

%\thib{
%\begin{itemize}
%    \item Here we focus on the upper gel phase
%    \item discuss first the impact of the carbon content; power-law scaling see Fig.~\ref{fig4}
%    \item discuss linear viscoelastic properties: Fractional Kelvin-Voigt model + rescaling of the viscoelastic properties; see Fig.~\ref{fig5}. 
%\end{itemize}
%}

We first report on the properties of gels obtained for $r>r_c$, whose microstructure is a viscoelastic matrix of CMC, physically cross-linked by CB nanoparticles dispersed individually or as clusters [see Fig.~\ref{fig:sketch_microstructure}(b)]. 

We first characterize the gel linear viscoelastic properties by determining its viscoelastic spectrum over three decades of frequencies $\omega$. Figure~\ref{fig6}(a)-(c) show the viscoelastic spectrum for $x_{\rm CB}=6, 8$ and 10\% wt.~at fixed CMC concentration $c_{\rm CMC}=2$\%. These three viscoelastic spectra all display a finite plateau modulus in the zero frequency limit, confirming that these mixtures behave as  soft solids at rest. Moreover, for these three spectra $G''(\omega)$ displays a power-law dependence, at least over two decades in frequency. Both behaviors are well captured by a \textit{fractional} Kelvin-Voigt model, which consists in a spring of stiffness $G_0$ in parallel with a spring-pot element \cite{Jaishankar:2013,Bonfanti:2020} characterized by a quasi-property $\mathbb{V}$ (dimension Pa.s$^\alpha$) and an exponent $0<\alpha<1$ [see inset in Fig.~\ref{fig6}(a) for a sketch of the mechanical model]. Such a fractional element, first introduced by Scott-Blair \cite{ScottBlair:1959} to describe the power-law spectrum of bitumen, yields the following expression for the complex modulus:
\begin{equation}
    G^\star = G_0 + \mathbb{V}(i\omega)^\alpha,
    \label{eq:FKV}
\end{equation}
which real and imaginary parts correspond to the red fits in Fig.~\ref{fig6}(a)--(c) determined simultaneously. This fit function relies only on two dimensional parameters, namely the elastic modulus $G_0$ and the characteristic frequency defined as $\omega_0=(G_0/\mathbb{V})^{1/\alpha}$. Remarkably, all  spectra obtained by varying the carbon content between 6 and 15\% wt.~at fixed $c_{\rm CMC}=2$\% can be described by varying $G_0$ and $\mathbb{V}$, while fixing $\alpha=2/3$. 

Such a robust description allows us to propose a universal master curve for the viscoelastic spectrum obtained at various CB contents, by using the following set of normalized coordinates: $\tilde G'=G'/G_0$, $\tilde G''=G''/G_0$, and $\tilde \omega = \omega/\omega_0$ [Fig.~\ref{fig6}(d)]. This master curve is, in turn, well described over 8 orders of magnitude of reduced frequency by the fractional Kelvin-Voigt model pictured as red lines in Fig.~\ref{fig6}(d). The power-law dependence of both $\tilde G'$ and $\tilde G''$ with an exponent $2/3$ in the high-frequency limit is in remarkable agreement with the value computed by Zimm for dense polymer suspensions and polymer solids taking into account Brownian motion and hydrodynamic interactions  \cite{Zimm:1956,Bagley:1983}. This observation strongly suggests that CB nanoparticles contribute mainly to the low-frequency part ($\tilde \omega \ll 1$) of the viscoelastic spectrum, whereas the CMC entangled network dominates the high-frequency response ($\tilde \omega \gg 1$). This result, together with the fact that CB-CMC hydrogels are not electrically conductive for $r>r_c$, provide robust evidence that in the regime $r>r_c$, the gel microstructure consists of a CMC viscoelastic matrix in which CB nanoparticles are dispersed without forming a percolated network, while serving as crosslinkers. Finally, note that the master curve is robust and holds for different CMC concentrations, as illustrated in supplemental Fig.~\ref{Sfig:master_KVF_CMC}.  

% In order to verify this interpretation, we have prepared CMC hydrogels in the absence of any CB nanoparticles. In that case, gel formation is triggered under acidic conditions by the decrease of the charge density of the polymer, which facilitates interchain associations \cite{Lopez:2021}. Here we use a gel made with $c_{\rm CMC}$~=~3\% and [HCl]~=~0.5~M, which yields pH~=~0.7. The viscoelastic spectrum of the gel indeed shows  a power-law dependence with an exponent $2/3$ in the high-frequency limit (see supplemental Fig.~\ref{Sfig:spectrum_CMC3_HCl0p5}). These results, together with the fact that CB-CMC hydrogels are not electrically conductive for $r>r_c$, provide robust evidence that in the regime $r>r_c$, the gel microstructure consists of a CMC viscoelastic matrix in which CB nanoparticles are dispersed without forming a percolated network, while serving as crosslinkers.  

\begin{figure}[!t]
    \centering
    \includegraphics[width=0.95\columnwidth]{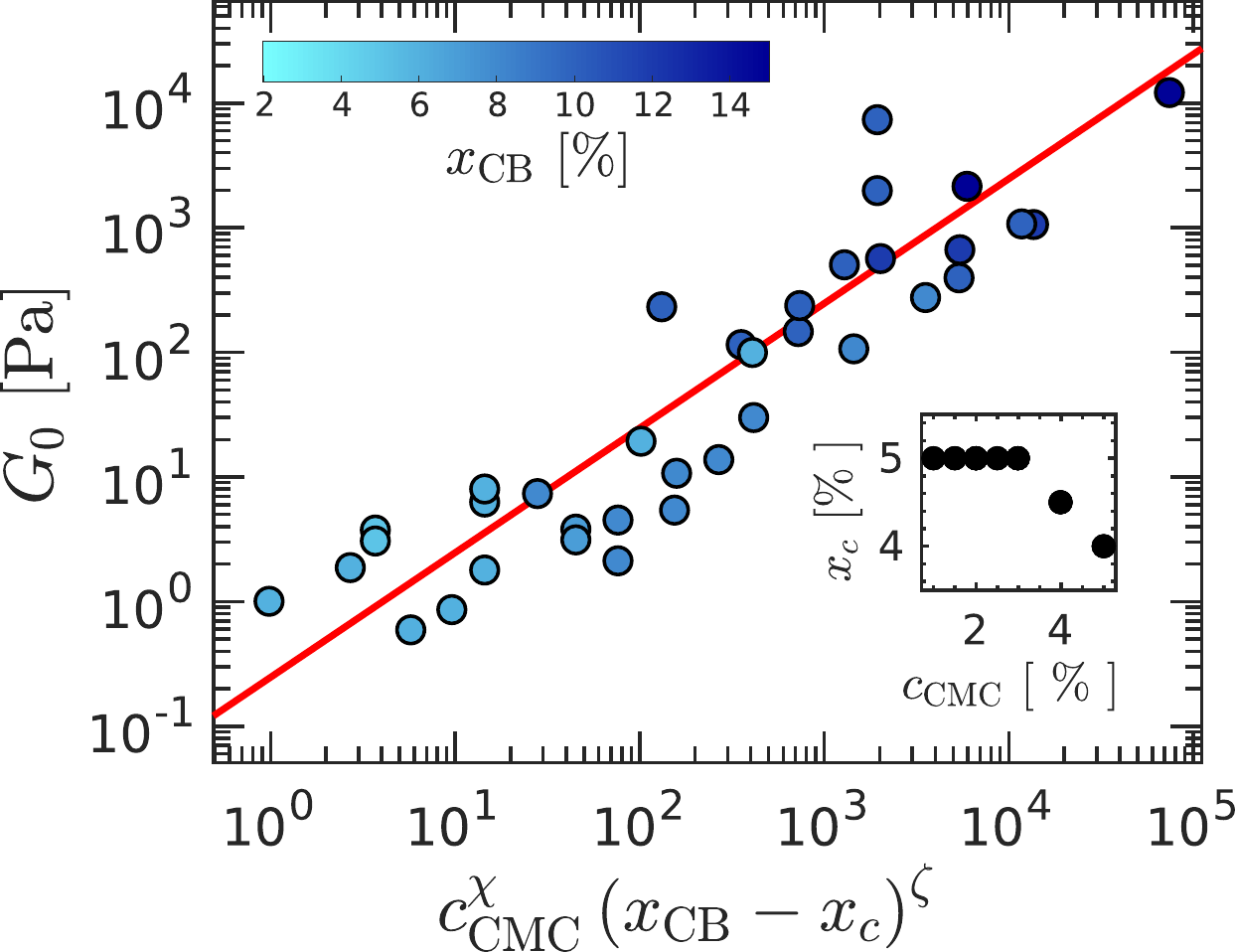}
    \caption{\textbf{Scaling of the elastic modulus in the polymer-dominated regime.} $G_0$ vs $c_{\rm CMC}^\chi (x_ {\rm CB} - x_c)^\zeta$ for $r>r_c$, where $\chi = 2.4 \pm 0.4$ and $\zeta = 3.0 \pm 0.5$. The inset shows the threshold of CB content $x_c$ as a function of $c_{\rm CMC}$ inferred from the sol-gel transition in the phase diagram reported in Fig.~\ref{fig2}. The red line shows the best linear fit of the data.}
    \label{G0_polym}
\end{figure}

\begin{figure*}[!h]
    \centering
    \includegraphics[width=1\textwidth]{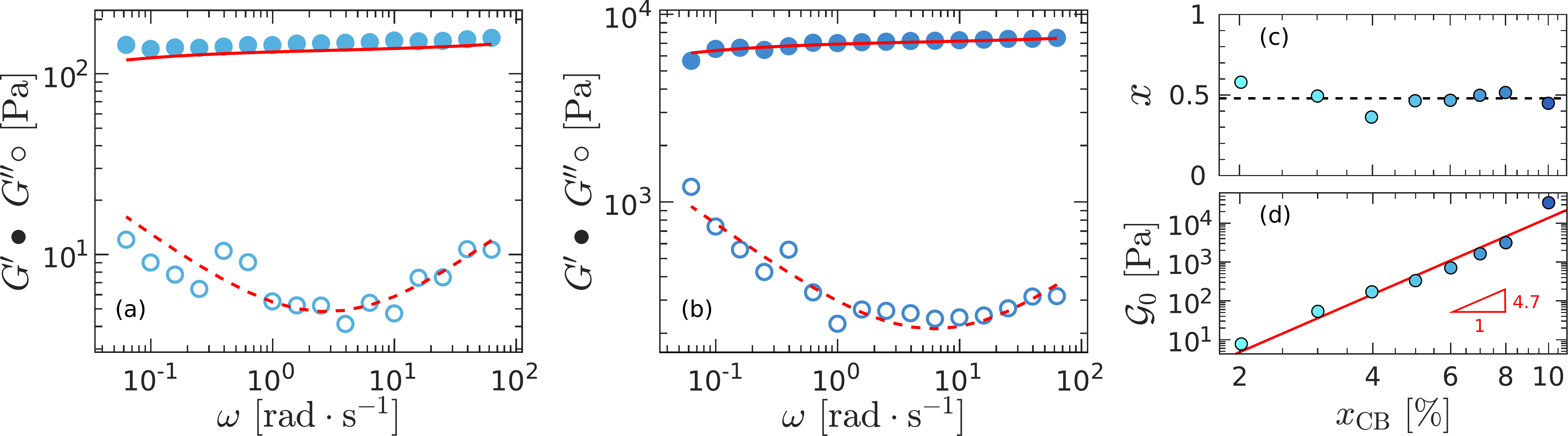}
    \caption{\textbf{Viscoelastic spectrum in the colloid-dominated regime.} Frequency dependence of the elastic and viscous moduli, $G'$ and $G''$ resp., of CMC-CB hydrogels with two different CB contents (a) $x_{\rm CB}=$ 6\%  and (b)  8\%~wt.~and a fixed amount of CMC polymer, i.e., $c_{\rm CMC}=10^{-2}$\%~wt. The red curves are the best fits of the data to a modified SGR model [see Eqs.~\eqref{eq:SGR1} and~\eqref{eq:SGR2}] with $\tau = $~1200~s, $x=0.50$, $c=1.4~\mathrm{Pa.s^{1/2}}$ in (a), and $\tau = $~1200~s, $x=0.53$, $c=41~\mathrm{Pa.s^{1/2}}$ in (b). (c) Fit parameter $x$, which is constant $x = 0.5 \pm 0.1$, irrespective of the CB content as shown by the dashed line. (d) Fit parameter $\mathcal{G}_0$, which increases as a power law of $x_{\rm CB}$ with an exponent $4.7 \pm 0.7$.}
    \label{fig7}
\end{figure*}

We now discuss the dependence of the plateau modulus $G_0$ that characterizes the gel elasticity as a function of both the CB content and the CMC concentration. As reported in Fig.~\ref{G0_polym}, all our experimental results collapse on a master curve when $G_0$ is plotted as a function of $c_{\rm CMC}^\chi (x_ {\rm CB} - x_c)^\zeta$. On the one hand, the gel elasticity increases as a power-law of the particle content relative to the mechanical percolation threshold $x_c$, which depends on the CMC concentration (see inset in Fig.~\ref{G0_polym}) and was determined from the phase diagram in Fig.~\ref{fig2}. In practice, the onset of rigidity decreases weakly for increasing polymer concentration, from $x_c=5$\% at $c_{\rm CMC}=1$\% to $x_c=4$\% at $c_{\rm CMC}=5$\%. On the other hand, the gel elasticity increases as a power law of the CMC concentration, with an exponent $\chi=2.4$. This scaling is strongly reminiscent of the concentration dependence predicted for the plateau modulus of entangled polymer solutions, for which an exponent $7/3\simeq 2.33$ was derived by Colby, Rubinstein and Viovy \cite{Colby:1992}. In the present case, for $r>r_c$, CMC are indeed in an entangled regime, which is reported\cite{Behra:2019} for $c_{\rm CMC} \gtrsim 0.16$\%~wt. This result further supports the idea that, for $r>r_c$, the elasticity of the CMC-CB hydrogel has two independent physical origins: a first contribution from the entangled CMC, and a second one from the CB particles, which serve as cross-linkers and inhibit the long-time relaxation of the CMC matrix, leading to solid-like behavior at rest.

\subsection{Colloid-dominated regime ($r<r_{\rm c}$)}

Let us now consider the case $r<r_c$, where the CMC-CB hydrogel microstructure is formed by a space-spanning network of CB nanoparticles decorated with CMC polymers [see Fig.~\ref{fig:sketch_microstructure}(a)]. The viscoelastic spectrum of two representative gels containing  $c_{\rm CMC} = 10^{-2} \%$ and $x_{\rm CB}=6$ and 8\% wt., respectively are illustrated in Fig.~\ref{fig7}(a) and (b). For both gels, $G'$ is merely frequency independent, whereas $G''$ decreases with increasing frequency, and shows a flattening or even a slight upturn in the high-frequency limit [see Fig.~\ref{fig7}(a)], which is the signature of the solvent viscosity \cite{Mason:1995}. Such a frequency spectrum, along with the logarithmic aging dynamics reported in supplemental Fig.~\ref{Sfig:time_sweep}(a)-(c), shows that CMC-CB hydrogels display a glassy-like behavior for $r<r_c$, similar to that reported for jammed assemblies of soft particles \cite{Purnomo:2008}, or fractal colloidal gels \cite{Aime:2018,Keshavarz:2021}. Therefore, we fit the viscoelastic spectrum reported in Fig.~\ref{fig7}(a) and \ref{fig7}(b) using a modified version of the Soft Glassy Rheology (SGR) model, which is well known to capture such a behavior and reads \cite{Sollich:1997,Sollich:1998,Fielding:2000,Purnomo:2006}:
\begin{align}
    G^{\prime} &= \mathcal{G}_0 \left(1 -  \left(\omega \tau \right)^{x-1}\right) + c \omega^{1/2} \label{eq:SGR1} \\
    G^{\prime \prime} &= \mathcal{G}_0 \left(\omega \tau\right)^{x-1} + c \omega^{1/2} + \eta_\infty \omega \label{eq:SGR2}
\end{align}
where $x$ corresponds to a mean-field noise temperature, $\tau$ denotes the effective sample age, and the terms $c\omega^{1/2}$ and $\eta_\infty \omega$ account respectively for Brownian motion, and for the solvent viscous contribution\cite{Mason:1995b}. The data are fitted using Eqs.~\eqref{eq:SGR1} and~\eqref{eq:SGR2} with $\mathcal{G}_0$, $x$, and $c$ as free parameters. The parameter $\tau = 1200$~s is the time elapsed since the end of the rejuvenation step and $\eta_{\infty} = 2.10^{-3}$~Pa.s is the viscosity of the CMC solution at $c_{\rm CMC} = 10^{-2} \%$, measured independently with an Ubbelholde viscosimeter. Fitting our experimental data for samples prepared with a CB content ranging between 2\% and 12\%, yields a constant effective temperature, $x\simeq 0.5<1$ [Fig.~\ref{fig7}(c)], that is characteristic of a yield stress fluid \cite{Fielding:2000}, and a reference modulus $\mathcal{G}_0$ that grows as a power-law of the CB weight fraction, with an exponent 4.7 [Fig.~\ref{fig7}(d)]. First, the latter value is much larger than 1.8, which allows us to rule out a microstructure involving polymer-bridged nanoparticles \cite{Surve:2006a,Surve:2006b}. Second, the exponent 4.7 is consistent with that reported for CB gels in aprotic solvents, e.g., mineral oils, in which the elasticity also originates from a percolated network of CB nanoparticles \cite{Trappe:2000,Grenard:2014}, as well as in depletion gels \cite{Prasad:2003}. Similarly, the Brownian parameter $c$ follows a power-law of the CB content as shown in Supplemental Fig.~\ref{Sfig:c_SGR}.

\begin{figure}[!t]
    \centering
    \includegraphics[width=0.95\columnwidth]{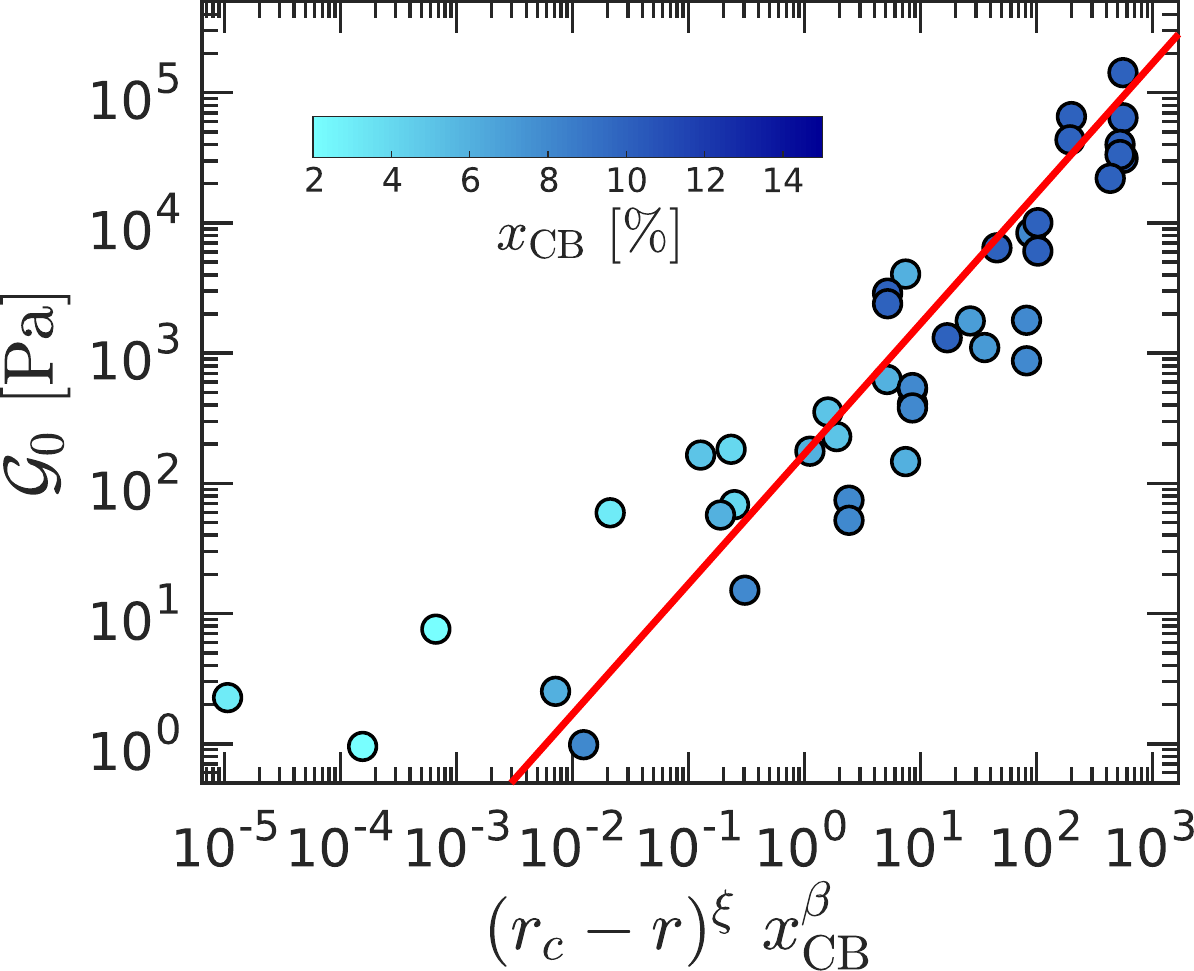}
    \caption{\textbf{Scaling of the elastic modulus in the colloid-dominated regime.} Elastic modulus $\mathcal{G}_0$ vs $(r_c -r)^\xi x_ {\rm CB} ^\beta$ for $r<r_c$, where $r_c = 0.037$, $\xi = 3.8 \pm 0.5$, and $\beta = 8.2 \pm 0.6$ (consistent with Fig.~\ref{fig4}). Color levels code for the CB content ranging between 2 and 10\%. The red line shows the best linearfit of the data.}
    \label{fig5}
\end{figure}

Varying the polymer concentration as well as the CB content, we observe that the gel reference modulus falls onto a master curve when plotted as a function of $(r_c -r)^\xi x_ {\rm CB} ^\beta$, as illustrated in Fig.~\ref{fig5} with $\xi = 3.8$ and $\beta = 8.2$. This expression accounts for the scalings displayed in Fig.~\ref{fig4} (a) and for the fact that, as $r\rightarrow r_c$, the elastic properties vanish, for the CB particles and clusters become isolated in the CMC matrix. Note as detailed in supplemental material, these exponents are fully consistent with the power-law dependence of $\mathcal{G}_0$ found in Fig.~\ref{fig7}.

\begin{figure}[t]
    \centering
    \includegraphics[width=0.95\columnwidth]{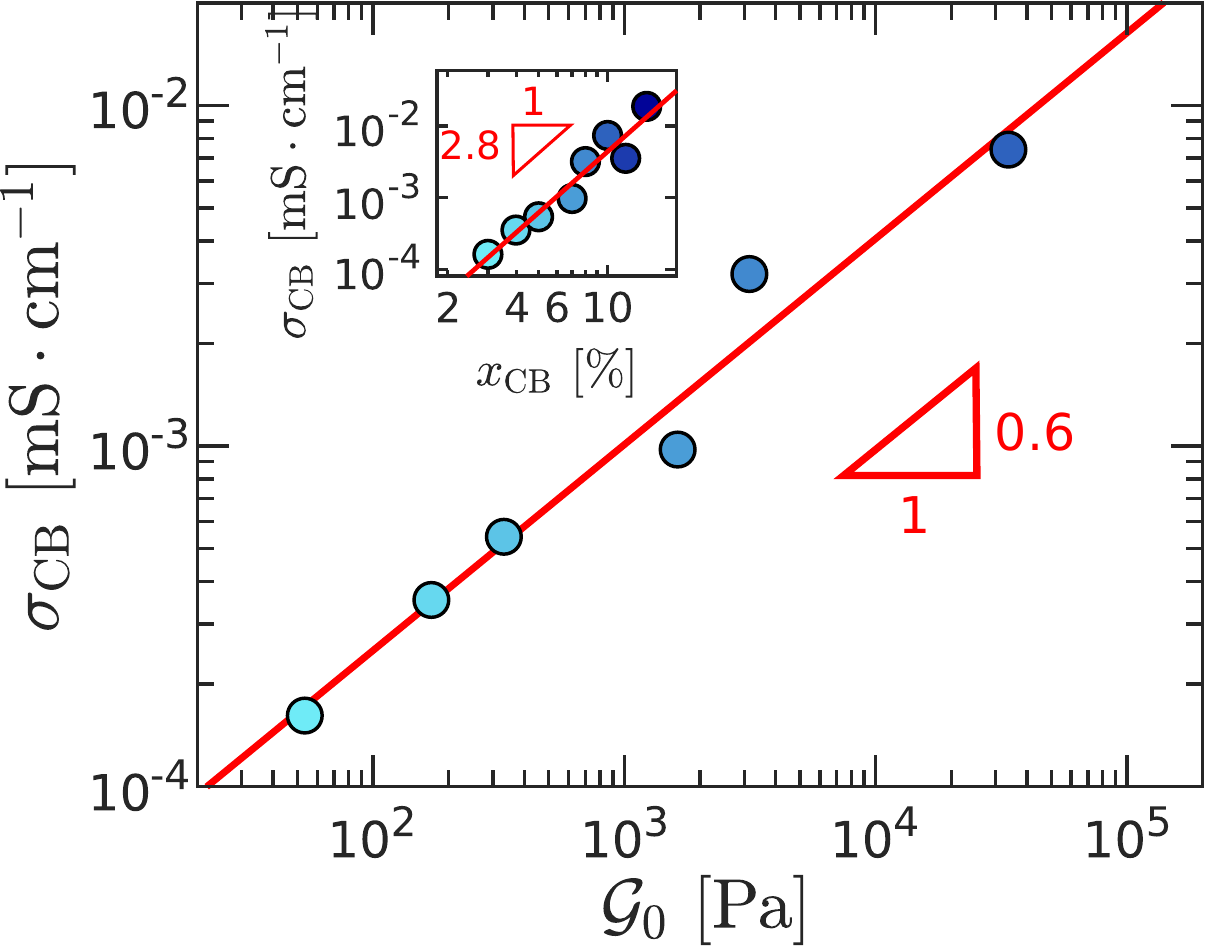}
    \caption{\textbf{Electrical conductivity of the CB network $\sigma_{\rm CB}$ vs the elastic modulus $G_0$.} The colors indicate the values of $x_{\rm CB}$ using the same scale as Fig.~\ref{fig5}. The red line is the best power-law fit with an exponent $0.5 \pm 0.2$. Inset: $\sigma_{\rm CB}$ vs $x_{\rm CB} - x_c$, with $x_c = 2 \%$ the percolation threshold determined in Fig.~\ref{fig2}. The red line is the best power-law fit with an exponent $1.8 \pm 0.4$.}
    \label{fig:sigma_vs_CB}
\end{figure}

Finally, CMC-CB hydrogels in the colloid-dominated regime (i.e., $r<r_c$) are electrically conductive [Fig.~\ref{fig4}(b)]. In Figure~\ref{fig:sigma_vs_CB}, we compare their mechanical and electrical properties, and observe that the electrical conductivity $\sigma_{\rm CB}$ conferred upon the sample by the percolated network of CB particles increases as a power law of the reference modulus $\mathcal{G}_0$, with an exponent of about 0.5. Note that a compatible exponent of 0.6 was robustly observed with the same type of CB nanoparticles dispersed in mineral oil at various weight fractions in the absence of any dispersant, and for various shear histories \cite{Helal:2016}. This comparison strongly suggests that, for $r<r_c$, CMC only serves to stabilize the CB network in water, without affecting the link between the mechanical and electrical properties, at least for $r \ll r_c$. Yet, one should emphasize that CMC-CB hydrogels display much lower conductivity values ranging between $10^{-4}$ and $10^{-2}$~mS/cm, compared to that measured for CB in mineral oil (between $10^{-2}$ and 2~mS/cm), at similar weight fractions. This discrepancy most likely results from the coating of the CB nanoparticles by the CMC, which lowers the conductivity.  

\section{Discussion and Conclusion}

By performing a fundamental and extensive study of aqueous dispersions of cellulose gum and carbon black soot particles, we have identified the range of polymer and nanoparticle contents for which the mixture forms hydrogels. These soft solids show two strikingly different microstructures and mechanical responses depending on the ratio $r$ that quantifies the relative content of polymers and nanoparticles.

For $r\ll r_c$, i.e., relatively low CMC concentrations, the gel elasticity is governed by the CB nanoparticles, which form a space-spanning network. Such a particulate network, which is responsible for the conductive properties of the hydrogel, is decorated and thus stabilized by the CMC. In this regime, the viscoelastic properties of the CMC-CB hydrogel, which are well-described by the SGR model, display weakly aging properties, which might be due to the slow reorganization of the particulate network. 

In contrast, for $r \gg r_c$, i.e., relatively high CMC concentrations, the gel elasticity originates from two independent contributions, namely the entangled CMC matrix on the one hand, and the CB nanoparticles on the other hand. In this regime, CB particles are dispersed inside the CMC matrix, where they act as crosslinkers through hydrophobic interactions. In consequence, the sample is not electronically conductive. Moreover, the viscoelastic spectrum of these CMC-CB hydrogels are frequency dependent and can be rescaled onto a master curve, which is very-well described by a fractional Kelvin-Voigt model with an exponent $\alpha=2/3$ that characterizes the high-frequency viscoelastic response. 

Interestingly, a similar master curve was reported for CMC-polydispersed silica hydrogels \cite{Pashkovski:2003}. Although the impact of the CMC concentration was not determined, the scaling factor $G_0$ was reported to increase as a power law of the silica content with an exponent 3.15 compatible with the value $\zeta=3.0\pm0.5$ reported in Fig.~\ref{G0_polym}. This comparison suggests that the nature and specific properties of the suspended nanoparticles, e.g., shape and size, have little or no influence on this exponent, which is mainly controlled by CMC. Moreover, the viscoelastic spectrum of CMC-silica hydrogels show a power-law in the high-frequency limit, with an exponent $0.56$ that is compatible with the Zimm scaling $\alpha=2/3$ reported in the present work. Note that this exponent is not universal, and depends on the polymer-solvent interaction. For instance, the viscoelastic spectrum of polyacrylamide-silica hydrogels obtained for various contents in silica particles \cite{Adibnia:2017} can also be rescaled onto a master curve that shows a high-frequency response that is a power law with an exponent $0.74$.

Our results pave the way for the rational design of CMC-based slurries in which CMC is used as a binder, e.g., for lithium-ion batteries \cite{Lee:2005,Dahbi:2014,Garcia:2018}, and more generally for CMC-based hydrogels involving various types of fillers such as graphene oxyde, metallic nanoparticles, etc. \cite{Bozaci:2015,Ali:2019,Saladino:2020}. Future work shall include rheo-electric studies of the yielding transition and flow properties of CMC-CB hydrogels, so as to unveil the microscopic scenario underpinning their non-linear properties.

\begin{acknowledgement}
The authors thank R.~Backov for suggesting the use of CMC as a dispersant, R.~Asmi for preliminary experiments, and C.~Bucher and V.~Andrieux for insightful discussions on the electrochemical properties of conductive fluids. We also acknowledge fruitful discussions with T.~Gibaud, A.~Helal, K.~Ioannidou, R.J.-M.~Pellenq, and M.~Peyla.

\end{acknowledgement}

%%%%%%%%%%%%%%%%%%%%%%%%%%%%%%%%%%%%%%%%%%%%%%%%%%%%%%%%%%%%%%%%%%%%%
%% The same is true for Supporting Information, which should use the
%% suppinfo environment.
%%%%%%%%%%%%%%%%%%%%%%%%%%%%%%%%%%%%%%%%%%%%%%%%%%%%%%%%%%%%%%%%%%%%%
\begin{suppinfo}

Supporting information include the ($i$) dynamic recovery of various CMC-CB hydrogels following flow cessation, ($ii$) calibration of the conductivity cell, ($iii$) phase diagrams based on the elastic and viscous modulus, ($iv$) scaling of the elastic modulus with the CMC concentration for $r<r_c$, ($v$) master curve for the hydrogel viscoelastic spectra for different CMC and CB contents, and ($iv$) dependence with the CB content of the parameter $c$ from the modified SGR model.

\end{suppinfo}

%%%%%%%%%%%%%%%%%%%%%%%%%%%%%%%%%%%%%%%%%%%%%%%%%%%%%%%%%%%%%%%%%%%%%
%% The appropriate \bibliography command should be placed here.
%% Notice that the class file automatically sets \bibliographystyle
%% and also names the section correctly.
%%%%%%%%%%%%%%%%%%%%%%%%%%%%%%%%%%%%%%%%%%%%%%%%%%%%%%%%%%%%%%%%%%%%%
%\bibliography{achemso-demo}

\begin{mcitethebibliography}{108}
\providecommand*\natexlab[1]{#1}
\providecommand*\mciteSetBstSublistMode[1]{}
\providecommand*\mciteSetBstMaxWidthForm[2]{}
\providecommand*\mciteBstWouldAddEndPuncttrue
  {\def\EndOfBibitem{\unskip.}}
\providecommand*\mciteBstWouldAddEndPunctfalse
  {\let\EndOfBibitem\relax}
\providecommand*\mciteSetBstMidEndSepPunct[3]{}
\providecommand*\mciteSetBstSublistLabelBeginEnd[3]{}
\providecommand*\EndOfBibitem{}
\mciteSetBstSublistMode{f}
\mciteSetBstMaxWidthForm{subitem}{(\alph{mcitesubitemcount})}
\mciteSetBstSublistLabelBeginEnd
  {\mcitemaxwidthsubitemform\space}
  {\relax}
  {\relax}

\bibitem[Samson \latin{et~al.}(1987)Samson, Mulholland, and
  Gentry]{Samson:1987}
Samson,~R.~J.; Mulholland,~G.~W.; Gentry,~J.~W. Structural analysis of soot
  agglomerates. \emph{Langmuir} \textbf{1987}, \emph{3}, 272--281\relax
\mciteBstWouldAddEndPuncttrue
\mciteSetBstMidEndSepPunct{\mcitedefaultmidpunct}
{\mcitedefaultendpunct}{\mcitedefaultseppunct}\relax
\EndOfBibitem{}

\bibitem[Spinelli(1998)]{Spinelli:1998}
Spinelli,~H.~J. Polymeric dispersants in ink jet technology. \emph{Adv. Mater.}
  \textbf{1998}, \emph{10}, 1215--1218\relax
\mciteBstWouldAddEndPuncttrue
\mciteSetBstMidEndSepPunct{\mcitedefaultmidpunct}
{\mcitedefaultendpunct}{\mcitedefaultseppunct}\relax
\EndOfBibitem{}

\bibitem[Ehrburger-Dolle \latin{et~al.}(2001)Ehrburger-Dolle,
  Hindermann-Bischoff, Livet, Bley, Rochas, and Geissler]{EhrburgerDolle:2001}
Ehrburger-Dolle,~F.; Hindermann-Bischoff,~M.; Livet,~F.; Bley,~F.; Rochas,~C.;
  Geissler,~E. Anisotropic Ultra-Small-Angle X-ray Scattering in Carbon Black
  Filled Polymers. \emph{Langmuir} \textbf{2001}, \emph{17}, 329--334\relax
\mciteBstWouldAddEndPuncttrue
\mciteSetBstMidEndSepPunct{\mcitedefaultmidpunct}
{\mcitedefaultendpunct}{\mcitedefaultseppunct}\relax
\EndOfBibitem{}

\bibitem[Huang(2002)]{Huang:2002}
Huang,~J.-C. Carbon black filled conducting polymers and polymer blends.
  \emph{Adv. Polym. Technol.} \textbf{2002}, \emph{21}, 299--313\relax
\mciteBstWouldAddEndPuncttrue
\mciteSetBstMidEndSepPunct{\mcitedefaultmidpunct}
{\mcitedefaultendpunct}{\mcitedefaultseppunct}\relax
\EndOfBibitem{}

\bibitem[Wen and Chung(2007)Wen, and Chung]{Wen:2007}
Wen,~S.; Chung,~D. D.~L. Partial replacement of carbon fiber by carbon black in
  multifunctional cement–matrix composites. \emph{Carbon} \textbf{2007},
  \emph{45}, 505--513\relax
\mciteBstWouldAddEndPuncttrue
\mciteSetBstMidEndSepPunct{\mcitedefaultmidpunct}
{\mcitedefaultendpunct}{\mcitedefaultseppunct}\relax
\EndOfBibitem{}

\bibitem[Pandolfo and Hollenkamp(2006)Pandolfo, and Hollenkamp]{Pandolfo:2006}
Pandolfo,~A.; Hollenkamp,~A.~F. Carbon properties and their role in
  supercapacitors. \emph{J. Power Sources} \textbf{2006}, \emph{157},
  11--27\relax
\mciteBstWouldAddEndPuncttrue
\mciteSetBstMidEndSepPunct{\mcitedefaultmidpunct}
{\mcitedefaultendpunct}{\mcitedefaultseppunct}\relax
\EndOfBibitem{}

\bibitem[Silva \latin{et~al.}(2017)Silva, Moraes, Janegitz, and
  Fatibello-Filho]{Silva:2017}
Silva,~T.~A.; Moraes,~F.~C.; Janegitz,~B.~C.; Fatibello-Filho,~O.
  Electrochemical biosensors based on nanostructured carbon black: A review.
  \emph{J. Nanomater.} \textbf{2017}, \emph{2017}\relax
\mciteBstWouldAddEndPuncttrue
\mciteSetBstMidEndSepPunct{\mcitedefaultmidpunct}
{\mcitedefaultendpunct}{\mcitedefaultseppunct}\relax
\EndOfBibitem{}

\bibitem[Kour \latin{et~al.}(2020)Kour, Arya, Young, Gupta, Bandhoria, and
  Khosla]{Kour:2020}
Kour,~R.; Arya,~S.; Young,~S.-J.; Gupta,~V.; Bandhoria,~P.; Khosla,~A. Recent
  advances in carbon nanomaterials as electrochemical biosensors. \emph{J.
  Electrochem. Soc.} \textbf{2020}, \emph{167}, 037555\relax
\mciteBstWouldAddEndPuncttrue
\mciteSetBstMidEndSepPunct{\mcitedefaultmidpunct}
{\mcitedefaultendpunct}{\mcitedefaultseppunct}\relax
\EndOfBibitem{}

\bibitem[Duduta \latin{et~al.}(2011)Duduta, Ho, Wood, Limthongkul, Brunini,
  Carter, and Chiang]{Duduta:2011}
Duduta,~M.; Ho,~B.; Wood,~V.~C.; Limthongkul,~P.; Brunini,~V.~E.;
  Carter,~W.~C.; Chiang,~Y.~M. Semi-Solid Lithium Rechargeable Flow Battery.
  \emph{Adv. Energy Mater.} \textbf{2011}, \emph{1}, 511--516\relax
\mciteBstWouldAddEndPuncttrue
\mciteSetBstMidEndSepPunct{\mcitedefaultmidpunct}
{\mcitedefaultendpunct}{\mcitedefaultseppunct}\relax
\EndOfBibitem{}

\bibitem[Li \latin{et~al.}(2013)Li, Smith, Dong, Baram, Fan, Xie, Limthongkul,
  Carter, and Chiang]{Li:2013b}
Li,~Z.; Smith,~K.~C.; Dong,~Y.; Baram,~N.; Fan,~F.~Y.; Xie,~J.;
  Limthongkul,~P.; Carter,~W.~C.; Chiang,~Y.-M. Aqueous semi-solid flow cell:
  demonstration and analysis. \emph{Phys. Chem. Chem. Phys.} \textbf{2013},
  \emph{15}, 15833--15839\relax
\mciteBstWouldAddEndPuncttrue
\mciteSetBstMidEndSepPunct{\mcitedefaultmidpunct}
{\mcitedefaultendpunct}{\mcitedefaultseppunct}\relax
\EndOfBibitem{}

\bibitem[Youssry \latin{et~al.}(2013)Youssry, Madec, Soudan, Cerbelaud,
  Guyomard, and Lestriez]{Youssry:2013}
Youssry,~M.; Madec,~L.; Soudan,~P.; Cerbelaud,~M.; Guyomard,~D.; Lestriez,~B.
  Non-aqueous carbon black suspensions for lithium-based redox flow batteries:
  rheology and simultaneous rheo-electrical behavior. \emph{Phys. Chem. Chem.
  Phys.} \textbf{2013}, \emph{15}, 14476--14486\relax
\mciteBstWouldAddEndPuncttrue
\mciteSetBstMidEndSepPunct{\mcitedefaultmidpunct}
{\mcitedefaultendpunct}{\mcitedefaultseppunct}\relax
\EndOfBibitem{}

\bibitem[Narayanan \latin{et~al.}(2017)Narayanan, Mugele, and
  Duits]{Narayanan:2017}
Narayanan,~A.; Mugele,~F.; Duits,~M. H.~G. Mechanical History Dependence in
  Carbon Black Suspensions for Flow Batteries: A Rheo-Impedance Study.
  \emph{Langmuir} \textbf{2017}, \emph{33}, 1629--1638\relax
\mciteBstWouldAddEndPuncttrue
\mciteSetBstMidEndSepPunct{\mcitedefaultmidpunct}
{\mcitedefaultendpunct}{\mcitedefaultseppunct}\relax
\EndOfBibitem{}

\bibitem[Narayanan \latin{et~al.}(2021)Narayanan, Zhu, Gen{\c{c}}er, McKinley,
  and Shao-Horn]{Narayanan:2021}
Narayanan,~T.~M.; Zhu,~Y.~G.; Gen{\c{c}}er,~E.; McKinley,~G.~H.; Shao-Horn,~Y.
  Low-cost manganese dioxide semi-solid electrode for flow batteries.
  \emph{Joule} \textbf{2021}, \emph{5}, 2934--2954\relax
\mciteBstWouldAddEndPuncttrue
\mciteSetBstMidEndSepPunct{\mcitedefaultmidpunct}
{\mcitedefaultendpunct}{\mcitedefaultseppunct}\relax
\EndOfBibitem{}

\bibitem[Khodabakhshi \latin{et~al.}(2020)Khodabakhshi, Fulvio, and
  Andreoli]{Khodabakhshi:2020}
Khodabakhshi,~S.; Fulvio,~P.~F.; Andreoli,~E. Carbon black reborn: Structure
  and chemistry for renewable energy harnessing. \emph{Carbon} \textbf{2020},
  \emph{162}, 604--649\relax
\mciteBstWouldAddEndPuncttrue
\mciteSetBstMidEndSepPunct{\mcitedefaultmidpunct}
{\mcitedefaultendpunct}{\mcitedefaultseppunct}\relax
\EndOfBibitem{}

\bibitem[Chou \latin{et~al.}(2014)Chou, Pan, Wang, Liu, and Dou]{Chou:2014}
Chou,~S.-L.; Pan,~Y.; Wang,~J.-Z.; Liu,~H.-K.; Dou,~S.-X. Small things make a
  big difference: binder effects on the performance of Li and Na batteries.
  \emph{Phys. Chem. Chem. Phys.} \textbf{2014}, \emph{16}, 20347--20359\relax
\mciteBstWouldAddEndPuncttrue
\mciteSetBstMidEndSepPunct{\mcitedefaultmidpunct}
{\mcitedefaultendpunct}{\mcitedefaultseppunct}\relax
\EndOfBibitem{}

\bibitem[Parant \latin{et~al.}(2017)Parant, Muller, {Le Mercier}, Tarascon,
  Poulin, and Colin]{Parant:2017}
Parant,~H.; Muller,~G.; {Le Mercier},~T.; Tarascon,~J.~M.; Poulin,~P.;
  Colin,~A. Flowing suspensions of carbon black with high electronic
  conductivity for flow applications: Comparison between carbons black and
  exhibition of specific aggregation of carbon particles. \emph{Carbon}
  \textbf{2017}, \emph{119}, 10--20\relax
\mciteBstWouldAddEndPuncttrue
\mciteSetBstMidEndSepPunct{\mcitedefaultmidpunct}
{\mcitedefaultendpunct}{\mcitedefaultseppunct}\relax
\EndOfBibitem{}

\bibitem[Youssry \latin{et~al.}(2018)Youssry, Kamand, Magzoub, and
  Nasser]{Youssry:2018}
Youssry,~M.; Kamand,~F.~Z.; Magzoub,~M.~I.; Nasser,~M.~S. Aqueous dispersions
  of carbon black and its hybrid with carbon nanofibers. \emph{RSC advances}
  \textbf{2018}, \emph{8}, 32119--32131\relax
\mciteBstWouldAddEndPuncttrue
\mciteSetBstMidEndSepPunct{\mcitedefaultmidpunct}
{\mcitedefaultendpunct}{\mcitedefaultseppunct}\relax
\EndOfBibitem{}

\bibitem[Meslam \latin{et~al.}(2022)Meslam, Elzatahry, and
  Youssry]{Meslam:2022}
Meslam,~M.; Elzatahry,~A.~A.; Youssry,~M. Promising aqueous dispersions of
  carbon black for semisolid flow battery application. \emph{Colloids Surf. A:
  Physicochem. Eng. Asp.} \textbf{2022}, 129376\relax
\mciteBstWouldAddEndPuncttrue
\mciteSetBstMidEndSepPunct{\mcitedefaultmidpunct}
{\mcitedefaultendpunct}{\mcitedefaultseppunct}\relax
\EndOfBibitem{}

\bibitem[van~der Waarden(1950)]{Waarden:1950}
van~der Waarden,~M. Stabilization of carbon black dispersion in hydrocarbons.
  \emph{J. Colloid Sci.} \textbf{1950}, \emph{5}, 317--325\relax
\mciteBstWouldAddEndPuncttrue
\mciteSetBstMidEndSepPunct{\mcitedefaultmidpunct}
{\mcitedefaultendpunct}{\mcitedefaultseppunct}\relax
\EndOfBibitem{}

\bibitem[Hartley and Parfitt(1985)Hartley, and Parfitt]{Hartley:1985}
Hartley,~P.~A.; Parfitt,~G.~D. Dispersion of powders in liquids. 1. The
  contribution of the van der Waals force to the cohesiveness of carbon black
  powders. \emph{Langmuir} \textbf{1985}, \emph{1}, 651--657\relax
\mciteBstWouldAddEndPuncttrue
\mciteSetBstMidEndSepPunct{\mcitedefaultmidpunct}
{\mcitedefaultendpunct}{\mcitedefaultseppunct}\relax
\EndOfBibitem{}

\bibitem[Trappe \latin{et~al.}(2007)Trappe, Pitard, Ramos, Robert, Bissig, and
  Cipelletti]{Trappe:2007}
Trappe,~V.; Pitard,~E.; Ramos,~L.; Robert,~A.; Bissig,~H.; Cipelletti,~L.
  Investigation of q-dependent dynamical heterogeneity in a colloidal gel by
  x-ray photon correlation spectroscopy. \emph{Phys. Rev. E} \textbf{2007},
  \emph{76}, 051404\relax
\mciteBstWouldAddEndPuncttrue
\mciteSetBstMidEndSepPunct{\mcitedefaultmidpunct}
{\mcitedefaultendpunct}{\mcitedefaultseppunct}\relax
\EndOfBibitem{}

\bibitem[Trappe and Weitz(2000)Trappe, and Weitz]{Trappe:2000}
Trappe,~V.; Weitz,~D.~A. Scaling of the viscoelasticity of weakly attractive
  particles. \emph{Phys. Rev. Lett.} \textbf{2000}, \emph{85}, 449--452\relax
\mciteBstWouldAddEndPuncttrue
\mciteSetBstMidEndSepPunct{\mcitedefaultmidpunct}
{\mcitedefaultendpunct}{\mcitedefaultseppunct}\relax
\EndOfBibitem{}

\bibitem[Trappe \latin{et~al.}(2001)Trappe, Prasad, Cipelletti, Segre, and
  Weitz]{Trappe:2001}
Trappe,~V.; Prasad,~V.; Cipelletti,~L.; Segre,~P.~N.; Weitz,~D.~A. Jamming
  phase diagram for attractive particles. \emph{Nature} \textbf{2001},
  \emph{411}, 772--775\relax
\mciteBstWouldAddEndPuncttrue
\mciteSetBstMidEndSepPunct{\mcitedefaultmidpunct}
{\mcitedefaultendpunct}{\mcitedefaultseppunct}\relax
\EndOfBibitem{}

\bibitem[Kawaguchi \latin{et~al.}(2001)Kawaguchi, Okuno, and
  Kato]{Kawaguchi:2001}
Kawaguchi,~M.; Okuno,~M.; Kato,~T. Rheological Properties of Carbon Black
  Suspensions in a Silicone Oil. \emph{Langmuir} \textbf{2001}, \emph{17},
  6041--6044\relax
\mciteBstWouldAddEndPuncttrue
\mciteSetBstMidEndSepPunct{\mcitedefaultmidpunct}
{\mcitedefaultendpunct}{\mcitedefaultseppunct}\relax
\EndOfBibitem{}

\bibitem[Gibaud \latin{et~al.}(2010)Gibaud, Frelat, and
  Manneville]{Gibaud:2010}
Gibaud,~T.; Frelat,~D.; Manneville,~S. Heterogeneous yielding dynamics in a
  colloidal gel. \emph{Soft Matter} \textbf{2010}, \emph{6}, 3482--3488\relax
\mciteBstWouldAddEndPuncttrue
\mciteSetBstMidEndSepPunct{\mcitedefaultmidpunct}
{\mcitedefaultendpunct}{\mcitedefaultseppunct}\relax
\EndOfBibitem{}

\bibitem[Sprakel \latin{et~al.}(2011)Sprakel, Lindstr\"om, Kodger, and
  Weitz]{Sprakel:2011}
Sprakel,~J.; Lindstr\"om,~S.; Kodger,~T.; Weitz,~D. Stress enhancement in the
  delayed yielding of colloidal gels. \emph{Phys. Rev. Lett.} \textbf{2011},
  \emph{106}, 248303\relax
\mciteBstWouldAddEndPuncttrue
\mciteSetBstMidEndSepPunct{\mcitedefaultmidpunct}
{\mcitedefaultendpunct}{\mcitedefaultseppunct}\relax
\EndOfBibitem{}

\bibitem[Grenard \latin{et~al.}(2014)Grenard, Divoux, Taberlet, and
  Manneville]{Grenard:2014}
Grenard,~V.; Divoux,~T.; Taberlet,~N.; Manneville,~S. Timescales in creep and
  yielding of attractive gels. \emph{Soft Matter} \textbf{2014}, \emph{10},
  1555--1571\relax
\mciteBstWouldAddEndPuncttrue
\mciteSetBstMidEndSepPunct{\mcitedefaultmidpunct}
{\mcitedefaultendpunct}{\mcitedefaultseppunct}\relax
\EndOfBibitem{}

\bibitem[Gibaud \latin{et~al.}(2016)Gibaud, Perge, Lindstr\"om, Taberlet, and
  Manneville]{Gibaud:2016}
Gibaud,~T.; Perge,~C.; Lindstr\"om,~S.~B.; Taberlet,~N.; Manneville,~S.
  Multiple yielding processes in a colloidal gel under large amplitude
  oscillatory stress. \emph{Soft Matter} \textbf{2016}, \emph{12},
  1701--1712\relax
\mciteBstWouldAddEndPuncttrue
\mciteSetBstMidEndSepPunct{\mcitedefaultmidpunct}
{\mcitedefaultendpunct}{\mcitedefaultseppunct}\relax
\EndOfBibitem{}

\bibitem[Ovarlez \latin{et~al.}(2013)Ovarlez, Tocquer, Bertrand, and
  Coussot]{Ovarlez:2013}
Ovarlez,~G.; Tocquer,~L.; Bertrand,~F.; Coussot,~P. Rheopexy and tunable yield
  stress of carbon black suspensions. \emph{Soft Matter} \textbf{2013},
  \emph{9}, 5540--5549\relax
\mciteBstWouldAddEndPuncttrue
\mciteSetBstMidEndSepPunct{\mcitedefaultmidpunct}
{\mcitedefaultendpunct}{\mcitedefaultseppunct}\relax
\EndOfBibitem{}

\bibitem[Divoux \latin{et~al.}(2013)Divoux, Grenard, and
  Manneville]{Divoux:2013}
Divoux,~T.; Grenard,~V.; Manneville,~S. Rheological hysteresis in soft glassy
  materials. \emph{Phys. Rev. Lett.} \textbf{2013}, \emph{110}, 018304\relax
\mciteBstWouldAddEndPuncttrue
\mciteSetBstMidEndSepPunct{\mcitedefaultmidpunct}
{\mcitedefaultendpunct}{\mcitedefaultseppunct}\relax
\EndOfBibitem{}

\bibitem[Helal \latin{et~al.}(2016)Helal, Divoux, and McKinley]{Helal:2016}
Helal,~H.; Divoux,~T.; McKinley,~G.~H. Simultaneous rheo-electric measurements
  of strongly conductive complex fluids. \emph{Phys. Rev. Applied}
  \textbf{2016}, \emph{6}, 064004\relax
\mciteBstWouldAddEndPuncttrue
\mciteSetBstMidEndSepPunct{\mcitedefaultmidpunct}
{\mcitedefaultendpunct}{\mcitedefaultseppunct}\relax
\EndOfBibitem{}

\bibitem[Li \latin{et~al.}(2005)Li, Chen, Xu, Yuan, Wang, and Wang]{Li:2005}
Li,~H.~Y.; Chen,~H.~Z.; Xu,~W.~J.; Yuan,~F.; Wang,~J.~R.; Wang,~M.
  Polymer-encapsulated hydrophilic carbon black nanoparticles free from
  aggregagtion. \emph{Colloid Surf. A: Physicocgem Eng. Aspect} \textbf{2005},
  \emph{254}, 173--178\relax
\mciteBstWouldAddEndPuncttrue
\mciteSetBstMidEndSepPunct{\mcitedefaultmidpunct}
{\mcitedefaultendpunct}{\mcitedefaultseppunct}\relax
\EndOfBibitem{}

\bibitem[Paredes \latin{et~al.}(2005)Paredes, Gracia, Mart{\'\i}nez-Alonso, and
  Tasc{\'o}n]{Paredes:2005}
Paredes,~J.; Gracia,~M.; Mart{\'\i}nez-Alonso,~A.; Tasc{\'o}n,~J. Nanoscale
  investigation of the structural and chemical changes induced by oxidation on
  carbon black surfaces: A scanning probe microscopy approach. \emph{J. Colloid
  Interface Sci.} \textbf{2005}, \emph{288}, 190--199\relax
\mciteBstWouldAddEndPuncttrue
\mciteSetBstMidEndSepPunct{\mcitedefaultmidpunct}
{\mcitedefaultendpunct}{\mcitedefaultseppunct}\relax
\EndOfBibitem{}

\bibitem[Liu \latin{et~al.}(2003)Liu, Jia, Kowalewski, Matyjaszewski,
  Casado-Portilla, and Belmont]{Liu:2003}
Liu,~T.; Jia,~S.; Kowalewski,~T.; Matyjaszewski,~K.; Casado-Portilla,~R.;
  Belmont,~J. Grafting poly (n-butyl acrylate) from a functionalized carbon
  black surface by atom transfer radical polymerization. \emph{Langmuir}
  \textbf{2003}, \emph{19}, 6342--6345\relax
\mciteBstWouldAddEndPuncttrue
\mciteSetBstMidEndSepPunct{\mcitedefaultmidpunct}
{\mcitedefaultendpunct}{\mcitedefaultseppunct}\relax
\EndOfBibitem{}

\bibitem[Liu \latin{et~al.}(2006)Liu, Jia, Kowalewski, Matyjaszewski,
  Casado-Portilla, and Belmont]{Liu:2006}
Liu,~T.; Jia,~S.; Kowalewski,~T.; Matyjaszewski,~K.; Casado-Portilla,~R.;
  Belmont,~J. Water-dispersible carbon black nanocomposites prepared by
  surface-initiated atom transfer radical polymerization in protic media.
  \emph{Macromolecules} \textbf{2006}, \emph{39}, 548--556\relax
\mciteBstWouldAddEndPuncttrue
\mciteSetBstMidEndSepPunct{\mcitedefaultmidpunct}
{\mcitedefaultendpunct}{\mcitedefaultseppunct}\relax
\EndOfBibitem{}

\bibitem[Lin \latin{et~al.}(1998)Lin, Chen, Wang, and Liaw]{Lin:1998}
Lin,~J.-H.; Chen,~H.-W.; Wang,~K.-T.; Liaw,~F.-H. A novel method for grafting
  polymers on carbon blacks. \emph{J. Mater. Chem.} \textbf{1998}, \emph{8},
  2169--2173\relax
\mciteBstWouldAddEndPuncttrue
\mciteSetBstMidEndSepPunct{\mcitedefaultmidpunct}
{\mcitedefaultendpunct}{\mcitedefaultseppunct}\relax
\EndOfBibitem{}

\bibitem[Tsubokawa(2002)]{Tsubokawa:2002}
Tsubokawa,~N. Functionalization of carbon material by surface grafting of
  polymers. \emph{Bull. Chem. Soc. Jpn.} \textbf{2002}, \emph{75},
  2115--2136\relax
\mciteBstWouldAddEndPuncttrue
\mciteSetBstMidEndSepPunct{\mcitedefaultmidpunct}
{\mcitedefaultendpunct}{\mcitedefaultseppunct}\relax
\EndOfBibitem{}

\bibitem[Yang \latin{et~al.}(2007)Yang, Wang, Xiang, Zhou, and
  Hua~Tan]{Yang:2007}
Yang,~Q.; Wang,~L.; Xiang,~W.; Zhou,~J.; Hua~Tan,~Q. A temperature-responsive
  carbon black nanoparticle prepared by surface-induced reversible
  addition--fragmentation chain transfer polymerization. \emph{Polymer}
  \textbf{2007}, \emph{48}, 3444--3451\relax
\mciteBstWouldAddEndPuncttrue
\mciteSetBstMidEndSepPunct{\mcitedefaultmidpunct}
{\mcitedefaultendpunct}{\mcitedefaultseppunct}\relax
\EndOfBibitem{}

\bibitem[Pei \latin{et~al.}(2014)Pei, Travas-Sejdic, and Williams]{Pei:2014}
Pei,~Y.; Travas-Sejdic,~J.; Williams,~D.~E. Water structure change-induced
  expansion and collapse of zwitterionic polymers surface-grafted onto carbon
  black. \emph{Aust. J. Chem.} \textbf{2014}, \emph{67}, 1706--1709\relax
\mciteBstWouldAddEndPuncttrue
\mciteSetBstMidEndSepPunct{\mcitedefaultmidpunct}
{\mcitedefaultendpunct}{\mcitedefaultseppunct}\relax
\EndOfBibitem{}

\bibitem[Wang \latin{et~al.}(2019)Wang, Zhang, Wang, Li, Du, and Fu]{Wang:2019}
Wang,~L.; Zhang,~L.; Wang,~D.; Li,~M.; Du,~C.; Fu,~S. Surface modification of
  carbon black by thiolene click reaction for improving dispersibility in
  aqueous phase. \emph{J. Dispers. Sci. Technol.} \textbf{2019}, \emph{40},
  152--160\relax
\mciteBstWouldAddEndPuncttrue
\mciteSetBstMidEndSepPunct{\mcitedefaultmidpunct}
{\mcitedefaultendpunct}{\mcitedefaultseppunct}\relax
\EndOfBibitem{}

\bibitem[Tiarks \latin{et~al.}(2001)Tiarks, Landfester, and
  Antonietti]{Tiarks:2001}
Tiarks,~F.; Landfester,~K.; Antonietti,~M. Encapsulation of carbon black by
  miniemulsion polymerization. \emph{Macromol. Chem. Phys.} \textbf{2001},
  \emph{202}, 51--60\relax
\mciteBstWouldAddEndPuncttrue
\mciteSetBstMidEndSepPunct{\mcitedefaultmidpunct}
{\mcitedefaultendpunct}{\mcitedefaultseppunct}\relax
\EndOfBibitem{}

\bibitem[Casado \latin{et~al.}(2007)Casado, Lovell, Navabpour, and
  Stanford]{Casado:2007}
Casado,~R.~M.; Lovell,~P.~A.; Navabpour,~P.; Stanford,~J.~L. Polymer
  encapsulation of surface-modified carbon blacks using surfactant-free
  emulsion polymerisation. \emph{Polymer} \textbf{2007}, \emph{48},
  2554--2563\relax
\mciteBstWouldAddEndPuncttrue
\mciteSetBstMidEndSepPunct{\mcitedefaultmidpunct}
{\mcitedefaultendpunct}{\mcitedefaultseppunct}\relax
\EndOfBibitem{}

\bibitem[Alves and Cooper(1981)Alves, and Cooper]{Alves:1981}
Alves,~B.~R.; Cooper,~W.~D. Colloid stabilisation by polyelectrolytes.
  Dispersions of carbon black in aqueous poly (acrylic acid) solution. \emph{J.
  Chem. Soc., Faraday Trans. 1} \textbf{1981}, \emph{77}, 889--896\relax
\mciteBstWouldAddEndPuncttrue
\mciteSetBstMidEndSepPunct{\mcitedefaultmidpunct}
{\mcitedefaultendpunct}{\mcitedefaultseppunct}\relax
\EndOfBibitem{}

\bibitem[Iijima \latin{et~al.}(2013)Iijima, Yamazaki, Nomura, and
  Kamiya]{Ijima:2013}
Iijima,~M.; Yamazaki,~M.; Nomura,~Y.; Kamiya,~H. Effect of structure of
  cationic dispersants on stability of carbon black nanoparticles and further
  processability through layer-by-layer surface modification. \emph{Chem. Eng.
  Sci.} \textbf{2013}, \emph{85}, 30--37\relax
\mciteBstWouldAddEndPuncttrue
\mciteSetBstMidEndSepPunct{\mcitedefaultmidpunct}
{\mcitedefaultendpunct}{\mcitedefaultseppunct}\relax
\EndOfBibitem{}

\bibitem[Hanada \latin{et~al.}(2013)Hanada, Masuda, Iijima, and
  Kamiya]{Hanada:2013}
Hanada,~Y.; Masuda,~S.; Iijima,~M.; Kamiya,~H. Analysis of dispersion and
  aggregation behavior of carbon black particles in aqueous suspension by
  colloid probe AFM method. \emph{Adv. Powder Technol.} \textbf{2013},
  \emph{24}, 844--851\relax
\mciteBstWouldAddEndPuncttrue
\mciteSetBstMidEndSepPunct{\mcitedefaultmidpunct}
{\mcitedefaultendpunct}{\mcitedefaultseppunct}\relax
\EndOfBibitem{}

\bibitem[Ogura \latin{et~al.}(1993)Ogura, Tanoura, and Hiraki]{Ogura:1993}
Ogura,~T.; Tanoura,~M.; Hiraki,~A. Behavior of surfactants in the suspension of
  coal components. \emph{Bull. Chem. Soc. Jpn.} \textbf{1993}, \emph{66},
  1633--1639\relax
\mciteBstWouldAddEndPuncttrue
\mciteSetBstMidEndSepPunct{\mcitedefaultmidpunct}
{\mcitedefaultendpunct}{\mcitedefaultseppunct}\relax
\EndOfBibitem{}

\bibitem[Ogura \latin{et~al.}(1994)Ogura, Tanoura, Tatsuhara, and
  Hiraki]{Ogura:1994}
Ogura,~T.; Tanoura,~M.; Tatsuhara,~K.; Hiraki,~A. The Role of Surfactants in
  Achieving Highly Loaded Aqueous Suspensions of Organic Particles. \emph{Bull.
  Chem. Soc. Jpn.} \textbf{1994}, \emph{67}, 3143--3149\relax
\mciteBstWouldAddEndPuncttrue
\mciteSetBstMidEndSepPunct{\mcitedefaultmidpunct}
{\mcitedefaultendpunct}{\mcitedefaultseppunct}\relax
\EndOfBibitem{}

\bibitem[Gupta and Bhagwat(2005)Gupta, and Bhagwat]{Gupta:2005}
Gupta,~S.~D.; Bhagwat,~S.~S. Adsorption of surfactants on carbon black-water
  interface. \emph{J. Dispers. Sci. Technol.} \textbf{2005}, \emph{26},
  111--120\relax
\mciteBstWouldAddEndPuncttrue
\mciteSetBstMidEndSepPunct{\mcitedefaultmidpunct}
{\mcitedefaultendpunct}{\mcitedefaultseppunct}\relax
\EndOfBibitem{}

\bibitem[Zhao \latin{et~al.}(2013)Zhao, Lu, Li, Fan, Wen, Zhan, Shu, Xu, and
  Zeng]{Zhao:2013}
Zhao,~Y.; Lu,~P.; Li,~C.; Fan,~X.; Wen,~Q.; Zhan,~Q.; Shu,~X.; Xu,~T.; Zeng,~G.
  Adsorption mechanism of sodium dodecyl benzene sulfonate on carbon blacks by
  adsorption isotherm and zeta potential determinations. \emph{Environ.
  Technol.} \textbf{2013}, \emph{34}, 201--207\relax
\mciteBstWouldAddEndPuncttrue
\mciteSetBstMidEndSepPunct{\mcitedefaultmidpunct}
{\mcitedefaultendpunct}{\mcitedefaultseppunct}\relax
\EndOfBibitem{}

\bibitem[Mizukawa and Kawaguchi(2009)Mizukawa, and Kawaguchi]{Mizukawa:2009}
Mizukawa,~H.; Kawaguchi,~M. Effects of perfluorosulfonic acid adsorption on the
  stability of carbon black suspensions. \emph{Langmuir} \textbf{2009},
  \emph{25}, 11984--11987\relax
\mciteBstWouldAddEndPuncttrue
\mciteSetBstMidEndSepPunct{\mcitedefaultmidpunct}
{\mcitedefaultendpunct}{\mcitedefaultseppunct}\relax
\EndOfBibitem{}

\bibitem[Subramanian and {\O}ye(2021)Subramanian, and {\O}ye]{Subramanian:2021}
Subramanian,~S.; {\O}ye,~G. Aqueous carbon black dispersions stabilized by
  sodium lignosulfonates. \emph{Colloid Polym. Sci.} \textbf{2021}, \emph{299},
  1223--1236\relax
\mciteBstWouldAddEndPuncttrue
\mciteSetBstMidEndSepPunct{\mcitedefaultmidpunct}
{\mcitedefaultendpunct}{\mcitedefaultseppunct}\relax
\EndOfBibitem{}

\bibitem[Bele \latin{et~al.}(1998)Bele, Kodre, Ar{\v{c}}on, Grdadolnik,
  Pejovnik, and Besenhard]{Bele:1998}
Bele,~M.; Kodre,~A.; Ar{\v{c}}on,~I.; Grdadolnik,~J.; Pejovnik,~S.;
  Besenhard,~J. Adsorption of cetyltrimethylammonium bromide on carbon black
  from aqueous solution. \emph{Carbon} \textbf{1998}, \emph{36},
  1207--1212\relax
\mciteBstWouldAddEndPuncttrue
\mciteSetBstMidEndSepPunct{\mcitedefaultmidpunct}
{\mcitedefaultendpunct}{\mcitedefaultseppunct}\relax
\EndOfBibitem{}

\bibitem[Porcher \latin{et~al.}(2010)Porcher, Lestriez, Jouanneau, and
  Guyomard]{Porcher:2010}
Porcher,~W.; Lestriez,~B.; Jouanneau,~S.; Guyomard,~D. Optimizing the
  surfactant for the aqueous processing of LiFePO4 composite electrodes.
  \emph{J. Power Sources} \textbf{2010}, \emph{195}, 2835--2843\relax
\mciteBstWouldAddEndPuncttrue
\mciteSetBstMidEndSepPunct{\mcitedefaultmidpunct}
{\mcitedefaultendpunct}{\mcitedefaultseppunct}\relax
\EndOfBibitem{}

\bibitem[Eisermann \latin{et~al.}(2014)Eisermann, Damm, Winzer, and
  Peukert]{Eisermann:2014}
Eisermann,~C.; Damm,~C.; Winzer,~B.; Peukert,~W. Stabilization of carbon black
  particles with cetyltrimethylammoniumbromide in aqueous media. \emph{Power
  Technology} \textbf{2014}, \emph{253}, 338--346\relax
\mciteBstWouldAddEndPuncttrue
\mciteSetBstMidEndSepPunct{\mcitedefaultmidpunct}
{\mcitedefaultendpunct}{\mcitedefaultseppunct}\relax
\EndOfBibitem{}

\bibitem[Ridaoui \latin{et~al.}(2006)Ridaoui, Jada, Vidal, and
  Donnet]{Ridaoui:2006}
Ridaoui,~H.; Jada,~A.; Vidal,~L.; Donnet,~J.-B. Effect of cationic surfactant
  and block copolymer on carbon black particle surface charge and size.
  \emph{Colloids Surf. A: Physicochem. Eng. Asp.} \textbf{2006}, \emph{278},
  149--159\relax
\mciteBstWouldAddEndPuncttrue
\mciteSetBstMidEndSepPunct{\mcitedefaultmidpunct}
{\mcitedefaultendpunct}{\mcitedefaultseppunct}\relax
\EndOfBibitem{}

\bibitem[Sis and Birinci(2009)Sis, and Birinci]{Sis:2009}
Sis,~H.; Birinci,~M. Effect of nonionic and ionic surfactants on zeta potential
  and dispersion properties of carbon black powders. \emph{Colloids Surf. A:
  Physicochem. Eng. Asp.} \textbf{2009}, \emph{341}, 60--67\relax
\mciteBstWouldAddEndPuncttrue
\mciteSetBstMidEndSepPunct{\mcitedefaultmidpunct}
{\mcitedefaultendpunct}{\mcitedefaultseppunct}\relax
\EndOfBibitem{}

\bibitem[Lin \latin{et~al.}(2002)Lin, Smith, and Alexandridis]{Lin:2002}
Lin,~Y.; Smith,~T.~W.; Alexandridis,~P. Adsorption of a rake-type siloxane
  surfactant onto carbon black nanoparticles dispersed in aqueous media.
  \emph{Langmuir} \textbf{2002}, \emph{18}, 6147--6158\relax
\mciteBstWouldAddEndPuncttrue
\mciteSetBstMidEndSepPunct{\mcitedefaultmidpunct}
{\mcitedefaultendpunct}{\mcitedefaultseppunct}\relax
\EndOfBibitem{}

\bibitem[Miano \latin{et~al.}(1992)Miano, Bailey, Luckham, and
  Tadros]{Miano:1992}
Miano,~F.; Bailey,~A.; Luckham,~P.; Tadros,~T.~F. Adsorption of nonyl phenol
  propylene oxide—ethylene oxide surfactants on carbon black and the rheology
  of the resulting dispersions. \emph{Colloids Surf.} \textbf{1992}, \emph{62},
  111--118\relax
\mciteBstWouldAddEndPuncttrue
\mciteSetBstMidEndSepPunct{\mcitedefaultmidpunct}
{\mcitedefaultendpunct}{\mcitedefaultseppunct}\relax
\EndOfBibitem{}

\bibitem[Yasin and Luckham(2012)Yasin, and Luckham]{Yasin:2012}
Yasin,~S.; Luckham,~P. Investigating the effectiveness of PEO/PPO based
  copolymers as dispersing agents for graphitic carbon black aqueous
  dispersions. \emph{Colloids Surf. A: Physicochem. Eng. Asp.} \textbf{2012},
  \emph{404}, 25--35\relax
\mciteBstWouldAddEndPuncttrue
\mciteSetBstMidEndSepPunct{\mcitedefaultmidpunct}
{\mcitedefaultendpunct}{\mcitedefaultseppunct}\relax
\EndOfBibitem{}

\bibitem[N’gouamba \latin{et~al.}(2020)N’gouamba, Goyon, Tocquer, Oerther,
  and Coussot]{Ngouamba:2020}
N’gouamba,~E.; Goyon,~J.; Tocquer,~L.; Oerther,~T.; Coussot,~P. Yielding,
  thixotropy, and strain stiffening of aqueous carbon black suspensions.
  \emph{J. Rheol.} \textbf{2020}, \emph{64}, 955--968\relax
\mciteBstWouldAddEndPuncttrue
\mciteSetBstMidEndSepPunct{\mcitedefaultmidpunct}
{\mcitedefaultendpunct}{\mcitedefaultseppunct}\relax
\EndOfBibitem{}

\bibitem[Guerfi \latin{et~al.}(2007)Guerfi, Kaneko, Petitclerc, Mori, and
  Zaghib]{Guerfi:2007}
Guerfi,~A.; Kaneko,~M.; Petitclerc,~M.; Mori,~M.; Zaghib,~K. LiFePO4
  water-soluble binder electrode for Li-ion batteries. \emph{J. Power Sources}
  \textbf{2007}, \emph{163}, 1047--1052\relax
\mciteBstWouldAddEndPuncttrue
\mciteSetBstMidEndSepPunct{\mcitedefaultmidpunct}
{\mcitedefaultendpunct}{\mcitedefaultseppunct}\relax
\EndOfBibitem{}

\bibitem[Lee \latin{et~al.}(2008)Lee, Kim, Kim, Zang, Choi, Park, and
  Paik]{Lee:2008}
Lee,~J.-H.; Kim,~J.-S.; Kim,~Y.~C.; Zang,~D.~S.; Choi,~Y.-M.; Park,~W.~I.;
  Paik,~U. Effect of carboxymethyl cellulose on aqueous processing of LiFePO4
  cathodes and their electrochemical performance. \emph{Electrochem.
  Solid-State Lett.} \textbf{2008}, \emph{11}, A175\relax
\mciteBstWouldAddEndPuncttrue
\mciteSetBstMidEndSepPunct{\mcitedefaultmidpunct}
{\mcitedefaultendpunct}{\mcitedefaultseppunct}\relax
\EndOfBibitem{}

\bibitem[Porcher \latin{et~al.}(2008)Porcher, Lestriez, Jouanneau, and
  Guyomard]{Porcher:2009}
Porcher,~W.; Lestriez,~B.; Jouanneau,~S.; Guyomard,~D. Design of aqueous
  processed thick LiFePO4 composite electrodes for high-energy lithium battery.
  \emph{J. Electrochem. Soc.} \textbf{2008}, \emph{156}, A133\relax
\mciteBstWouldAddEndPuncttrue
\mciteSetBstMidEndSepPunct{\mcitedefaultmidpunct}
{\mcitedefaultendpunct}{\mcitedefaultseppunct}\relax
\EndOfBibitem{}

\bibitem[Garc{\'\i}a \latin{et~al.}(2018)Garc{\'\i}a, Culebras, Collins, and
  Leahy]{Garcia:2018}
Garc{\'\i}a,~A.; Culebras,~M.; Collins,~M.~N.; Leahy,~J.~J. Stability and
  rheological study of sodium carboxymethyl cellulose and alginate suspensions
  as binders for lithium ion batteries. \emph{J. Appl. Polym. Sci.}
  \textbf{2018}, \emph{135}, 46217\relax
\mciteBstWouldAddEndPuncttrue
\mciteSetBstMidEndSepPunct{\mcitedefaultmidpunct}
{\mcitedefaultendpunct}{\mcitedefaultseppunct}\relax
\EndOfBibitem{}

\bibitem[Barrie \latin{et~al.}(2004)Barrie, Griffiths, Abbott, Grillo,
  Kudryashov, and Smyth]{Barrie:2004}
Barrie,~C.; Griffiths,~P.; Abbott,~R.; Grillo,~I.; Kudryashov,~E.; Smyth,~C.
  Rheology of aqueous carbon black dispersions. \emph{J. Colloid Interface
  Sci.} \textbf{2004}, \emph{272}, 210--217\relax
\mciteBstWouldAddEndPuncttrue
\mciteSetBstMidEndSepPunct{\mcitedefaultmidpunct}
{\mcitedefaultendpunct}{\mcitedefaultseppunct}\relax
\EndOfBibitem{}

\bibitem[Aoki \latin{et~al.}(2003)Aoki, Hatano, and Watanabe]{Aoki:2003}
Aoki,~Y.; Hatano,~A.; Watanabe,~H. Rheology of carbon black suspensions. I.
  Three types of viscoelastic behavior. \emph{Rheol. Acta} \textbf{2003},
  \emph{42}, 209--216\relax
\mciteBstWouldAddEndPuncttrue
\mciteSetBstMidEndSepPunct{\mcitedefaultmidpunct}
{\mcitedefaultendpunct}{\mcitedefaultseppunct}\relax
\EndOfBibitem{}

\bibitem[Rahman \latin{et~al.}(2021)Rahman, Hasan, Nitai, Nam, Karmakar, Ahsan,
  Shiddiky, Ahmed, \latin{et~al.} others]{Rahman:2021}
Rahman,~M.; Hasan,~M.; Nitai,~A.~S.; Nam,~S.; Karmakar,~A.~K.; Ahsan,~M.;
  Shiddiky,~M.~J.; Ahmed,~M.~B., \latin{et~al.}  Recent developments of
  carboxymethyl cellulose. \emph{Polymers} \textbf{2021}, \emph{13}, 1345\relax
\mciteBstWouldAddEndPuncttrue
\mciteSetBstMidEndSepPunct{\mcitedefaultmidpunct}
{\mcitedefaultendpunct}{\mcitedefaultseppunct}\relax
\EndOfBibitem{}

\bibitem[Jiang \latin{et~al.}(2009)Jiang, Li, Zhang, and Wang]{Jiang:2009}
Jiang,~L.; Li,~Y.; Zhang,~L.; Wang,~X. Preparation and characterization of a
  novel composite containing carboxymethyl cellulose used for bone repair.
  \emph{Mater. Sci. Eng. C} \textbf{2009}, \emph{29}, 193--198\relax
\mciteBstWouldAddEndPuncttrue
\mciteSetBstMidEndSepPunct{\mcitedefaultmidpunct}
{\mcitedefaultendpunct}{\mcitedefaultseppunct}\relax
\EndOfBibitem{}

\bibitem[Kulicke \latin{et~al.}(1996)Kulicke, Kull, Kull, Thielking,
  Engelhardt, and Pannek]{Kulicke:1996}
Kulicke,~W.-M.; Kull,~A.~H.; Kull,~W.; Thielking,~H.; Engelhardt,~J.;
  Pannek,~J.-B. Characterization of aqueous carboxymethylcellulose solutions in
  terms of their molecular structure and its influence on rheological
  behaviour. \emph{Polymer} \textbf{1996}, \emph{37}, 2723--2731\relax
\mciteBstWouldAddEndPuncttrue
\mciteSetBstMidEndSepPunct{\mcitedefaultmidpunct}
{\mcitedefaultendpunct}{\mcitedefaultseppunct}\relax
\EndOfBibitem{}

\bibitem[Clasen and Kulicke(2001)Clasen, and Kulicke]{Clasen:2001}
Clasen,~C.; Kulicke,~W.-M. Determination of viscoelastic and rheo-optical
  material functions of water-soluble cellulose derivatives. \emph{Prog. Polym.
  Sci.} \textbf{2001}, \emph{26}, 1839--1919\relax
\mciteBstWouldAddEndPuncttrue
\mciteSetBstMidEndSepPunct{\mcitedefaultmidpunct}
{\mcitedefaultendpunct}{\mcitedefaultseppunct}\relax
\EndOfBibitem{}

\bibitem[Ghannam and Esmail(1997)Ghannam, and Esmail]{Ghannam:1997}
Ghannam,~M.~T.; Esmail,~M.~N. Rheological properties of carboxymethyl
  cellulose. \emph{J. Appl. Polym. Sci.} \textbf{1997}, \emph{64},
  289--301\relax
\mciteBstWouldAddEndPuncttrue
\mciteSetBstMidEndSepPunct{\mcitedefaultmidpunct}
{\mcitedefaultendpunct}{\mcitedefaultseppunct}\relax
\EndOfBibitem{}

\bibitem[Benchabane and Bekkour(2008)Benchabane, and Bekkour]{Benchabane:2008}
Benchabane,~A.; Bekkour,~K. Rheological properties of carboxymethyl cellulose
  (CMC) solutions. \emph{Colloid Polym. Sci.} \textbf{2008}, \emph{286},
  1173--1180\relax
\mciteBstWouldAddEndPuncttrue
\mciteSetBstMidEndSepPunct{\mcitedefaultmidpunct}
{\mcitedefaultendpunct}{\mcitedefaultseppunct}\relax
\EndOfBibitem{}

\bibitem[Elliott and Ganz(1971)Elliott, and Ganz]{Elliott:1971}
Elliott,~J.~H.; Ganz,~A.~J. Modification of food characteristics with cellulose
  hydrocolloids I: Rheological characterization of an organoleptic property
  (unctuousness). \emph{J. Texture Stud.} \textbf{1971}, \emph{2},
  220--229\relax
\mciteBstWouldAddEndPuncttrue
\mciteSetBstMidEndSepPunct{\mcitedefaultmidpunct}
{\mcitedefaultendpunct}{\mcitedefaultseppunct}\relax
\EndOfBibitem{}

\bibitem[Barba \latin{et~al.}(2002)Barba, Montan{\'e}, Farriol, Desbri{\`e}res,
  and Rinaudo]{Barba:2002}
Barba,~C.; Montan{\'e},~D.; Farriol,~X.; Desbri{\`e}res,~J.; Rinaudo,~M.
  Synthesis and characterization of carboxymethylcelluloses from non-wood pulps
  II. Rheological behavior of CMC in aqueous solution. \emph{Cellulose}
  \textbf{2002}, \emph{9}, 327--335\relax
\mciteBstWouldAddEndPuncttrue
\mciteSetBstMidEndSepPunct{\mcitedefaultmidpunct}
{\mcitedefaultendpunct}{\mcitedefaultseppunct}\relax
\EndOfBibitem{}

\bibitem[Medronho \latin{et~al.}(2012)Medronho, Romano, Miguel, Stigsson, and
  Lindman]{Medronho:2012}
Medronho,~B.; Romano,~A.; Miguel,~M.~G.; Stigsson,~L.; Lindman,~B.
  Rationalizing cellulose (in) solubility: reviewing basic physicochemical
  aspects and role of hydrophobic interactions. \emph{Cellulose} \textbf{2012},
  \emph{19}, 581--587\relax
\mciteBstWouldAddEndPuncttrue
\mciteSetBstMidEndSepPunct{\mcitedefaultmidpunct}
{\mcitedefaultendpunct}{\mcitedefaultseppunct}\relax
\EndOfBibitem{}

\bibitem[Glasser \latin{et~al.}(2012)Glasser, Atalla, Blackwell, Malcolm~Brown,
  Burchard, French, Klemm, and Nishiyama]{Glasser:2012}
Glasser,~W.~G.; Atalla,~R.~H.; Blackwell,~J.; Malcolm~Brown,~R.; Burchard,~W.;
  French,~A.~D.; Klemm,~D.~O.; Nishiyama,~Y. About the structure of cellulose:
  debating the Lindman hypothesis. \emph{Cellulose} \textbf{2012}, \emph{19},
  589--598\relax
\mciteBstWouldAddEndPuncttrue
\mciteSetBstMidEndSepPunct{\mcitedefaultmidpunct}
{\mcitedefaultendpunct}{\mcitedefaultseppunct}\relax
\EndOfBibitem{}

\bibitem[Lopez \latin{et~al.}(2018)Lopez, Colby, and Cabral]{Lopez:2018}
Lopez,~C.~G.; Colby,~R.~H.; Cabral,~J.~T. Electrostatic and hydrophobic
  interactions in NaCMC aqueous solutions: Effect of degree of substitution.
  \emph{Macromolecules} \textbf{2018}, \emph{51}, 3165--3175\relax
\mciteBstWouldAddEndPuncttrue
\mciteSetBstMidEndSepPunct{\mcitedefaultmidpunct}
{\mcitedefaultendpunct}{\mcitedefaultseppunct}\relax
\EndOfBibitem{}

\bibitem[Lopez and Richtering(2021)Lopez, and Richtering]{Lopez:2021}
Lopez,~C.~G.; Richtering,~W. Oscillatory rheology of carboxymethyl cellulose
  gels: Influence of concentration and pH. \emph{Carbohydr. Polym.}
  \textbf{2021}, \emph{267}, 118117\relax
\mciteBstWouldAddEndPuncttrue
\mciteSetBstMidEndSepPunct{\mcitedefaultmidpunct}
{\mcitedefaultendpunct}{\mcitedefaultseppunct}\relax
\EndOfBibitem{}

\bibitem[Cole(1928)]{Cole:1928}
Cole,~K.~S. Electric impedance of suspensions of spheres. \emph{The Journal of
  general physiology} \textbf{1928}, \emph{12}, 29\relax
\mciteBstWouldAddEndPuncttrue
\mciteSetBstMidEndSepPunct{\mcitedefaultmidpunct}
{\mcitedefaultendpunct}{\mcitedefaultseppunct}\relax
\EndOfBibitem{}

\bibitem[Macdonald and Barsoukov(2018)Macdonald, and Barsoukov]{Macdonald:2018}
Macdonald,~J.~R.; Barsoukov,~E. \emph{Impedance spectroscopy: theory,
  experiment, and applications}; John Wiley \& Sons, 2018\relax
\mciteBstWouldAddEndPuncttrue
\mciteSetBstMidEndSepPunct{\mcitedefaultmidpunct}
{\mcitedefaultendpunct}{\mcitedefaultseppunct}\relax
\EndOfBibitem{}

\bibitem[Lasia(2022)]{Lasia:2022}
Lasia,~A. The origin of the constant phase element. \emph{J. Phys. Chem. Lett.}
  \textbf{2022}, \emph{13}, 580--589\relax
\mciteBstWouldAddEndPuncttrue
\mciteSetBstMidEndSepPunct{\mcitedefaultmidpunct}
{\mcitedefaultendpunct}{\mcitedefaultseppunct}\relax
\EndOfBibitem{}

\bibitem[Jaishankar and McKinley(2013)Jaishankar, and
  McKinley]{Jaishankar:2013}
Jaishankar,~A.; McKinley,~G.~H. Power-law rheology in the bulk and at the
  interface: quasi-properties and fractional constitutive equations.
  \emph{Proc. R. Soc. A: Math. Phys. Eng. Sci.} \textbf{2013}, \emph{469},
  20120284\relax
\mciteBstWouldAddEndPuncttrue
\mciteSetBstMidEndSepPunct{\mcitedefaultmidpunct}
{\mcitedefaultendpunct}{\mcitedefaultseppunct}\relax
\EndOfBibitem{}

\bibitem[Bonfanti \latin{et~al.}(2020)Bonfanti, Kaplan, Charras, and
  Kabla]{Bonfanti:2020}
Bonfanti,~A.; Kaplan,~J.~L.; Charras,~G.; Kabla,~A. Fractional viscoelastic
  models for power-law materials. \emph{Soft Matter} \textbf{2020}, \emph{16},
  6002--6020\relax
\mciteBstWouldAddEndPuncttrue
\mciteSetBstMidEndSepPunct{\mcitedefaultmidpunct}
{\mcitedefaultendpunct}{\mcitedefaultseppunct}\relax
\EndOfBibitem{}

\bibitem[{Scott-Blair} and Burnett(1959){Scott-Blair}, and
  Burnett]{ScottBlair:1959}
{Scott-Blair},~G.~W.; Burnett,~J. On the creep, recovery, relaxation and
  elastic ``memory" of some renneted milk gels. \emph{Br. J. Appl. Phys.}
  \textbf{1959}, \emph{10}, 15\relax
\mciteBstWouldAddEndPuncttrue
\mciteSetBstMidEndSepPunct{\mcitedefaultmidpunct}
{\mcitedefaultendpunct}{\mcitedefaultseppunct}\relax
\EndOfBibitem{}

\bibitem[Zimm(1956)]{Zimm:1956}
Zimm,~B.~H. Dynamics of polymer molecules in dilute solution: viscoelasticity,
  flow birefringence and dielectric loss. \emph{J. Chem. Phys.} \textbf{1956},
  \emph{24}, 269--278\relax
\mciteBstWouldAddEndPuncttrue
\mciteSetBstMidEndSepPunct{\mcitedefaultmidpunct}
{\mcitedefaultendpunct}{\mcitedefaultseppunct}\relax
\EndOfBibitem{}

\bibitem[Bagley and Torvik(1983)Bagley, and Torvik]{Bagley:1983}
Bagley,~R.~L.; Torvik,~P. A theoretical basis for the application of fractional
  calculus to viscoelasticity. \emph{J. Rheol.} \textbf{1983}, \emph{27},
  201--210\relax
\mciteBstWouldAddEndPuncttrue
\mciteSetBstMidEndSepPunct{\mcitedefaultmidpunct}
{\mcitedefaultendpunct}{\mcitedefaultseppunct}\relax
\EndOfBibitem{}

\bibitem[Colby \latin{et~al.}(1992)Colby, Rubinstein, and Viovy]{Colby:1992}
Colby,~R.~H.; Rubinstein,~M.; Viovy,~J.~L. Chain entanglement in polymer melts
  and solutions. \emph{Macromolecules} \textbf{1992}, \emph{25}, 996--998\relax
\mciteBstWouldAddEndPuncttrue
\mciteSetBstMidEndSepPunct{\mcitedefaultmidpunct}
{\mcitedefaultendpunct}{\mcitedefaultseppunct}\relax
\EndOfBibitem{}

\bibitem[Behra \latin{et~al.}(2019)Behra, Mattsson, Cayre, Robles, Tang, and
  Hunter]{Behra:2019}
Behra,~J.~S.; Mattsson,~J.; Cayre,~O.~J.; Robles,~E.~S.; Tang,~H.;
  Hunter,~T.~N. Characterization of sodium carboxymethyl cellulose aqueous
  solutions to support complex product formulation: A rheology and light
  scattering study. \emph{ACS Appl. Mater. Interfaces.} \textbf{2019},
  \emph{1}, 344--358\relax
\mciteBstWouldAddEndPuncttrue
\mciteSetBstMidEndSepPunct{\mcitedefaultmidpunct}
{\mcitedefaultendpunct}{\mcitedefaultseppunct}\relax
\EndOfBibitem{}

\bibitem[Mason and Weitz(1995)Mason, and Weitz]{Mason:1995}
Mason,~T.~G.; Weitz,~D.~A. Optical measurements of frequency-dependent linear
  viscoelastic moduli of complex fluids. \emph{Phys. Rev. Lett.} \textbf{1995},
  \emph{74}, 1250\relax
\mciteBstWouldAddEndPuncttrue
\mciteSetBstMidEndSepPunct{\mcitedefaultmidpunct}
{\mcitedefaultendpunct}{\mcitedefaultseppunct}\relax
\EndOfBibitem{}

\bibitem[Purnomo \latin{et~al.}(2008)Purnomo, Van Den~Ende, Vanapalli, and
  Mugele]{Purnomo:2008}
Purnomo,~E.~H.; Van Den~Ende,~D.; Vanapalli,~S.~A.; Mugele,~F. Glass transition
  and aging in dense suspensions of thermosensitive microgel particles.
  \emph{Phys. Rev. Lett.} \textbf{2008}, \emph{101}, 238301\relax
\mciteBstWouldAddEndPuncttrue
\mciteSetBstMidEndSepPunct{\mcitedefaultmidpunct}
{\mcitedefaultendpunct}{\mcitedefaultseppunct}\relax
\EndOfBibitem{}

\bibitem[Aime \latin{et~al.}(2018)Aime, Cipelletti, and Ramos]{Aime:2018}
Aime,~S.; Cipelletti,~L.; Ramos,~L. Power law viscoelasticity of a fractal
  colloidal gel. \emph{J. Rheol.} \textbf{2018}, \emph{62}, 1429--1441\relax
\mciteBstWouldAddEndPuncttrue
\mciteSetBstMidEndSepPunct{\mcitedefaultmidpunct}
{\mcitedefaultendpunct}{\mcitedefaultseppunct}\relax
\EndOfBibitem{}

\bibitem[Keshavarz \latin{et~al.}(2021)Keshavarz, Rodrigues, Champenois, Frith,
  Ilavsky, Geri, Divoux, McKinley, and Poulesquen]{Keshavarz:2021}
Keshavarz,~B.; Rodrigues,~D.~G.; Champenois,~J.-B.; Frith,~M.~G.; Ilavsky,~J.;
  Geri,~M.; Divoux,~T.; McKinley,~G.~H.; Poulesquen,~A. Time--connectivity
  superposition and the gel/glass duality of weak colloidal gels. \emph{Proc.
  Natl. Acad. Sci. U.S.A.} \textbf{2021}, \emph{118}, e2022339118\relax
\mciteBstWouldAddEndPuncttrue
\mciteSetBstMidEndSepPunct{\mcitedefaultmidpunct}
{\mcitedefaultendpunct}{\mcitedefaultseppunct}\relax
\EndOfBibitem{}

\bibitem[Sollich \latin{et~al.}(1997)Sollich, Lequeux, H{\'e}braud, and
  Cates]{Sollich:1997}
Sollich,~P.; Lequeux,~F.; H{\'e}braud,~P.; Cates,~M.~E. Rheology of soft glassy
  materials. \emph{Phys. Rev. Lett.} \textbf{1997}, \emph{78}, 2020\relax
\mciteBstWouldAddEndPuncttrue
\mciteSetBstMidEndSepPunct{\mcitedefaultmidpunct}
{\mcitedefaultendpunct}{\mcitedefaultseppunct}\relax
\EndOfBibitem{}

\bibitem[Sollich(1998)]{Sollich:1998}
Sollich,~P. Rheological constitutive equation for a model of soft glassy
  materials. \emph{Phys. Rev. E} \textbf{1998}, \emph{58}, 738--759\relax
\mciteBstWouldAddEndPuncttrue
\mciteSetBstMidEndSepPunct{\mcitedefaultmidpunct}
{\mcitedefaultendpunct}{\mcitedefaultseppunct}\relax
\EndOfBibitem{}

\bibitem[Fielding \latin{et~al.}(2000)Fielding, Sollich, and
  Cates]{Fielding:2000}
Fielding,~S.~M.; Sollich,~P.; Cates,~M.~E. Aging and rheology in soft
  materials. \emph{J. Rheol.} \textbf{2000}, \emph{44}, 323--369\relax
\mciteBstWouldAddEndPuncttrue
\mciteSetBstMidEndSepPunct{\mcitedefaultmidpunct}
{\mcitedefaultendpunct}{\mcitedefaultseppunct}\relax
\EndOfBibitem{}

\bibitem[Purnomo \latin{et~al.}(2006)Purnomo, Van Den~Ende, Mellema, and
  Mugele]{Purnomo:2006}
Purnomo,~E.; Van Den~Ende,~D.; Mellema,~J.; Mugele,~F. Linear viscoelastic
  properties of aging suspensions. \emph{Europhys. Lett.} \textbf{2006},
  \emph{76}, 74\relax
\mciteBstWouldAddEndPuncttrue
\mciteSetBstMidEndSepPunct{\mcitedefaultmidpunct}
{\mcitedefaultendpunct}{\mcitedefaultseppunct}\relax
\EndOfBibitem{}

\bibitem[Mason and Weitz(1995)Mason, and Weitz]{Mason:1995b}
Mason,~T.; Weitz,~D. Linear viscoelasticity of colloidal hard sphere
  suspensions near the glass transition. \emph{Phys. Rev. Lett.} \textbf{1995},
  \emph{75}, 2770\relax
\mciteBstWouldAddEndPuncttrue
\mciteSetBstMidEndSepPunct{\mcitedefaultmidpunct}
{\mcitedefaultendpunct}{\mcitedefaultseppunct}\relax
\EndOfBibitem{}

\bibitem[Surve \latin{et~al.}(2006)Surve, Pryamitsyn, and Ganesan]{Surve:2006a}
Surve,~M.; Pryamitsyn,~V.; Ganesan,~V. Universality in structure and elasticity
  of polymer-nanoparticle gels. \emph{Phys. Rev. Lett.} \textbf{2006},
  \emph{96}, 177805\relax
\mciteBstWouldAddEndPuncttrue
\mciteSetBstMidEndSepPunct{\mcitedefaultmidpunct}
{\mcitedefaultendpunct}{\mcitedefaultseppunct}\relax
\EndOfBibitem{}

\bibitem[Surve \latin{et~al.}(2006)Surve, Pryamitsyn, and Ganesan]{Surve:2006b}
Surve,~M.; Pryamitsyn,~V.; Ganesan,~V. Polymer-bridged gels of nanoparticles in
  solutions of adsorbing polymers. \emph{J. Chem. Phys.} \textbf{2006},
  \emph{125}, 064903\relax
\mciteBstWouldAddEndPuncttrue
\mciteSetBstMidEndSepPunct{\mcitedefaultmidpunct}
{\mcitedefaultendpunct}{\mcitedefaultseppunct}\relax
\EndOfBibitem{}

\bibitem[Prasad \latin{et~al.}(2003)Prasad, Trappe, Dinsmore, Segre,
  Cipelletti, and Weitz]{Prasad:2003}
Prasad,~V.; Trappe,~V.; Dinsmore,~A.~D.; Segre,~P.~N.; Cipelletti,~L.;
  Weitz,~D.~A. Universal features of the fluid to solid transition for
  attractive colloidal particles. \emph{Faraday Discuss.} \textbf{2003},
  \emph{123}, 1--12\relax
\mciteBstWouldAddEndPuncttrue
\mciteSetBstMidEndSepPunct{\mcitedefaultmidpunct}
{\mcitedefaultendpunct}{\mcitedefaultseppunct}\relax
\EndOfBibitem{}

\bibitem[Pashkovski \latin{et~al.}(2003)Pashkovski, Masters, and
  Mehreteab]{Pashkovski:2003}
Pashkovski,~E.~E.; Masters,~J.~G.; Mehreteab,~A. Viscoelastic scaling of
  colloidal gels in polymer solutions. \emph{Langmuir} \textbf{2003},
  \emph{19}, 3589--3595\relax
\mciteBstWouldAddEndPuncttrue
\mciteSetBstMidEndSepPunct{\mcitedefaultmidpunct}
{\mcitedefaultendpunct}{\mcitedefaultseppunct}\relax
\EndOfBibitem{}

\bibitem[Adibnia and Hill(2017)Adibnia, and Hill]{Adibnia:2017}
Adibnia,~V.; Hill,~R.~J. Viscoelasticity of near-critical silica-polyacrylamide
  hydrogel nanocomposites. \emph{Polymer} \textbf{2017}, \emph{112},
  457--465\relax
\mciteBstWouldAddEndPuncttrue
\mciteSetBstMidEndSepPunct{\mcitedefaultmidpunct}
{\mcitedefaultendpunct}{\mcitedefaultseppunct}\relax
\EndOfBibitem{}

\bibitem[Lee \latin{et~al.}(2005)Lee, Paik, Hackley, and Choi]{Lee:2005}
Lee,~J.-H.; Paik,~U.; Hackley,~V.~A.; Choi,~Y.-M. Effect of carboxymethyl
  cellulose on aqueous processing of natural graphite negative electrodes and
  their electrochemical performance for lithium batteries. \emph{J.
  Electrochem. Soc.} \textbf{2005}, \emph{152}, A1763\relax
\mciteBstWouldAddEndPuncttrue
\mciteSetBstMidEndSepPunct{\mcitedefaultmidpunct}
{\mcitedefaultendpunct}{\mcitedefaultseppunct}\relax
\EndOfBibitem{}

\bibitem[Dahbi \latin{et~al.}(2014)Dahbi, Nakano, Yabuuchi, Ishikawa, Kubota,
  Fukunishi, Shibahara, Son, Cui, Oji, \latin{et~al.} others]{Dahbi:2014}
Dahbi,~M.; Nakano,~T.; Yabuuchi,~N.; Ishikawa,~T.; Kubota,~K.; Fukunishi,~M.;
  Shibahara,~S.; Son,~J.-Y.; Cui,~Y.-T.; Oji,~H., \latin{et~al.}  Sodium
  carboxymethyl cellulose as a potential binder for hard-carbon negative
  electrodes in sodium-ion batteries. \emph{Electrochem. commun.}
  \textbf{2014}, \emph{44}, 66--69\relax
\mciteBstWouldAddEndPuncttrue
\mciteSetBstMidEndSepPunct{\mcitedefaultmidpunct}
{\mcitedefaultendpunct}{\mcitedefaultseppunct}\relax
\EndOfBibitem{}

\bibitem[Bozaci \latin{et~al.}(2015)Bozaci, Akar, Ozdogan, Demir, Altinisik,
  and Seki]{Bozaci:2015}
Bozaci,~E.; Akar,~E.; Ozdogan,~E.; Demir,~A.; Altinisik,~A.; Seki,~Y.
  Application of carboxymethylcellulose hydrogel based silver nanocomposites on
  cotton fabrics for antibacterial property. \emph{Carbohydr. Polym.}
  \textbf{2015}, \emph{134}, 128--135\relax
\mciteBstWouldAddEndPuncttrue
\mciteSetBstMidEndSepPunct{\mcitedefaultmidpunct}
{\mcitedefaultendpunct}{\mcitedefaultseppunct}\relax
\EndOfBibitem{}

\bibitem[Ali \latin{et~al.}(2019)Ali, Amin, and Ng]{Ali:2019}
Ali,~N.~H.; Amin,~M. C. I.~M.; Ng,~S.-F. Sodium carboxymethyl cellulose
  hydrogels containing reduced graphene oxide (rGO) as a functional antibiofilm
  wound dressing. \emph{J. Biomater. Sci. Polym. Ed.} \textbf{2019}, \emph{30},
  629--645\relax
\mciteBstWouldAddEndPuncttrue
\mciteSetBstMidEndSepPunct{\mcitedefaultmidpunct}
{\mcitedefaultendpunct}{\mcitedefaultseppunct}\relax
\EndOfBibitem{}

\bibitem[Saladino \latin{et~al.}(2020)Saladino, Markowska, Carmone, Cancemi,
  Alduina, Presentato, Scaffaro, Bia{\l}y, Hasiak, Hreniak, \latin{et~al.}
  others]{Saladino:2020}
Saladino,~M.~L.; Markowska,~M.; Carmone,~C.; Cancemi,~P.; Alduina,~R.;
  Presentato,~A.; Scaffaro,~R.; Bia{\l}y,~D.; Hasiak,~M.; Hreniak,~D.,
  \latin{et~al.}  Graphene oxide carboxymethylcellulose nanocomposite for
  dressing materials. \emph{Materials} \textbf{2020}, \emph{13}, 1980\relax
\mciteBstWouldAddEndPuncttrue
\mciteSetBstMidEndSepPunct{\mcitedefaultmidpunct}
{\mcitedefaultendpunct}{\mcitedefaultseppunct}\relax
\EndOfBibitem{}

\end{mcitethebibliography}

\providecommand{\latin}[1]{#1}
\makeatletter
\providecommand{\doi}
  {\begingroup\let\do\@makeother\dospecials
  \catcode`\{=1 \catcode`\}=2 \doi@aux}
\providecommand{\doi@aux}[1]{\endgroup\texttt{#1}}
\makeatother
\providecommand*\mcitethebibliography{\thebibliography}
\csname @ifundefined\endcsname{endmcitethebibliography}
  {\let\endmcitethebibliography\endthebibliography}{}

\clearpage
\newpage
\onecolumn
\setcounter{page}{1}
\setcounter{equation}{0}
\setcounter{figure}{0}
\global\def\thefigure{S\arabic{figure}}
\setcounter{table}{0}
\global\def\thetable{S\arabic{table}}

\begin{center}
    {\large\bf {\sc Supplementary information}}
\end{center}
\begin{center}
    {\large\bf Dual origin of viscoelasticity in polymer-carbon black hydrogels:\\ a rheometry and electrical spectroscopy study}
\end{center}

\section{Dynamics of recovery of CMC-CB hydrogels}
\begin{figure*}[!h]
    \centering
    \includegraphics[width=0.95\linewidth]{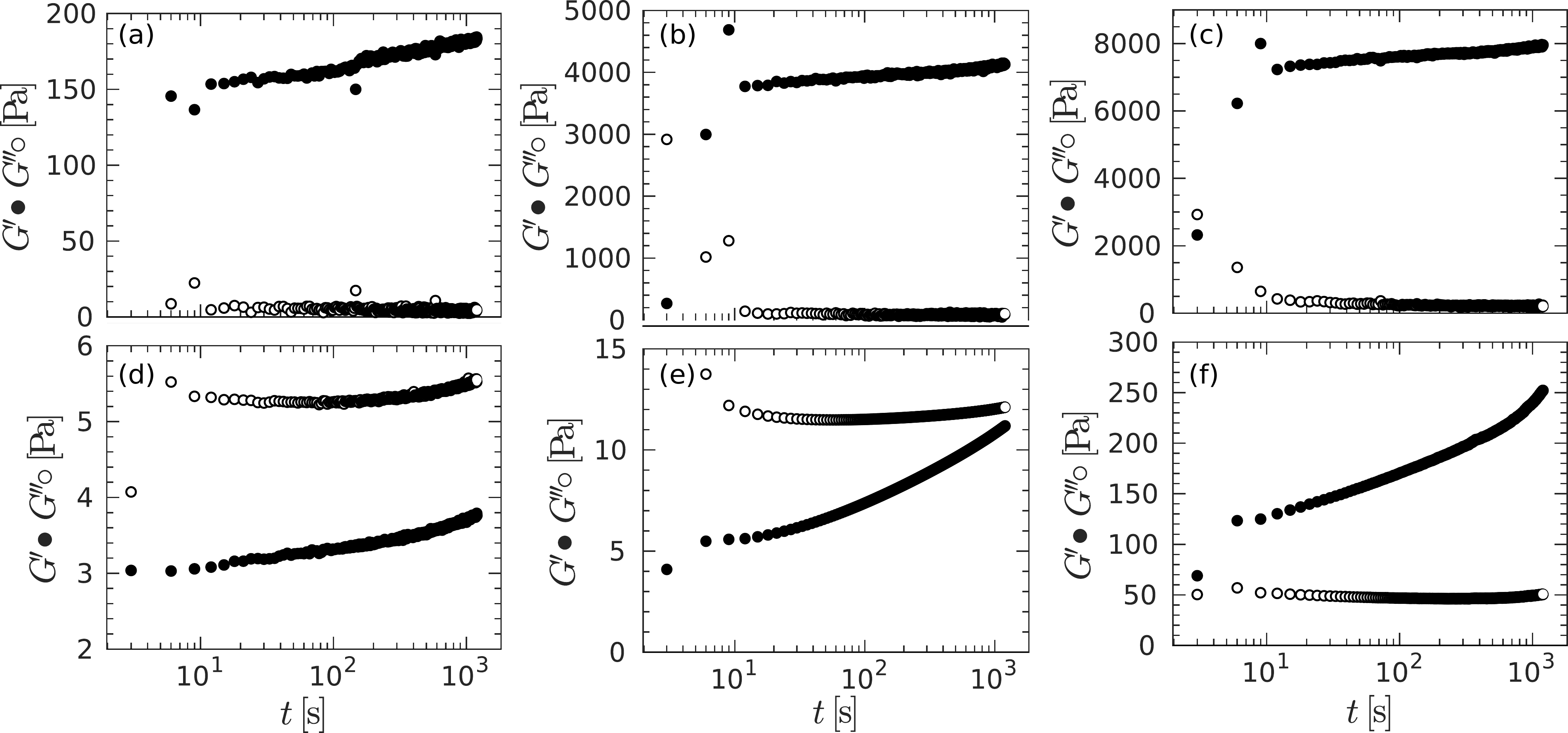}
    \caption{Recovery of CMC-CB hydrogels following a preshear at $\dot{\gamma} = 500 ~\rm s^{-1}$ during $180$: time evolution of the elastic and viscous moduli $G^{\prime}$ and $G^{\prime \prime}$ resp., measured at $f= 1~\rm Hz$ for two different CMC concentrations: $c_{\rm CMC} = 10^{-2} \%$ with (a)  $x_{\rm CB} = 4 \% $, (b) $x_{\rm CB} = 6 \% $, (c) $x_{\rm CB} = 8 \% $, and $c_{\rm CMC} = 2 \%$ with (d)  $x_{\rm CB} = 6 \% $, (e) $x_{\rm CB} = 8 \% $, (f)  $x_{\rm CB} = 10 \% $.}
    \label{Sfig:time_sweep}
\end{figure*}

Following a preshear at $\dot{\gamma} = 500 ~\rm s^{-1}$ during $180$~s, CMC-CB hydrogels display two different types of recoveries depending on the relative content of CMC and CB, as illustrated in Fig.~\ref{Sfig:time_sweep}. For $r < r_c$, the sample elastic modulus verifies $G'\gg G''$ in less than a few seconds, and grows logarithmically as a function of time [Fig.~\ref{Sfig:time_sweep}(a)-(c)]. Such dynamics are robust and merely depend on the CB content. In contrast, for $r > r_c$, the recovery is faster than logarithmic, and strongly depends on the CB content, i.e., the crossover time of $G'$ and $G''$ drops for increasing CB content [Fig.~\ref{Sfig:time_sweep}(d)-(f)].\\

\newpage

\section{Calibration of the conductivity cell}
\begin{figure}[!h]
    \centering
    \includegraphics[width=0.5\linewidth]{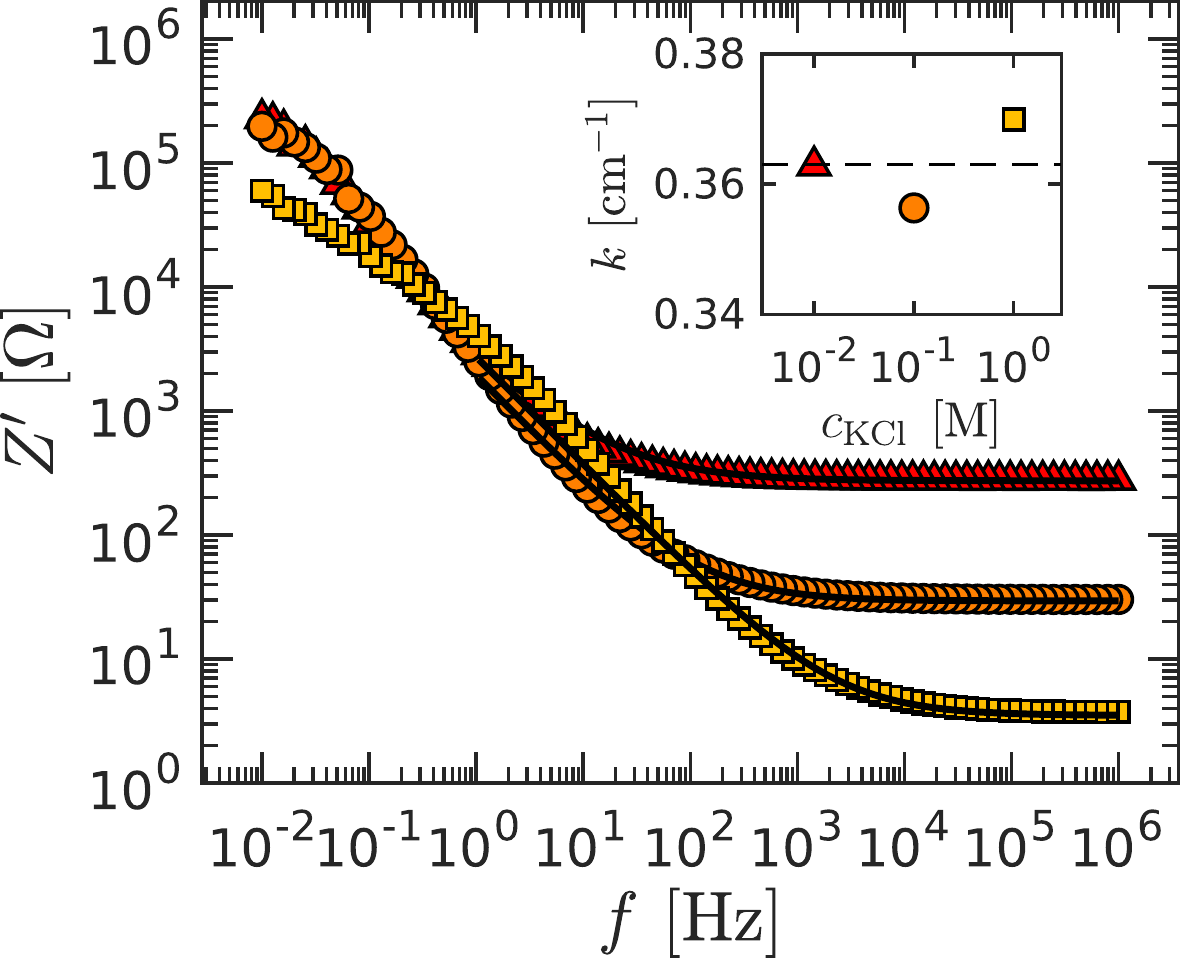}
    \caption{Real part (resistance) of the electrical impedance vs frequency of KCl solutions as measured with the conductivity cell for three different KCl concentrations: $c_{\rm KCl} = 10^{-2}$~M ($\blacktriangle$), $10^{-1}$~M ($\bullet$) and $1$~M ($\blacksquare$). The black lines are the best fits of the data with Eq.~\eqref{eq:elec2}. Inset: cell constant $k$ vs. KCl concentration, yielding $k = 0.36 \pm 0.01 ~ \rm cm^{-1}$.}
    \label{Sfig:calibration}
\end{figure}

The actual dimensions of the cell are unknown and the cell requires to be calibrated using KCl solutions of known concentration and conductivity (Certified conductivity OIML Standard Solution from Radiometer Analytical). Measurements were performed at room temperature at $T=22\pm1^\circ$C. The electrical impedance of these solutions is reported in Fig.~\ref{Sfig:calibration}, and fitted by the following equation:
\begin{equation}
    Z^*(\omega)=R_{\rm ion} + \frac{1}{Q(i\omega)^n}
\label{eq:elec2}
\end{equation}
which is equivalent to Eq.~\eqref{eq:elec} in the main document with $R_{\rm CB} = \infty$ (no CB network). The cell constant $k$ is then computed to match the expected ionic conductivity $\sigma_{\rm ion}$ at the given temperature and concentration in KCl so that $k = R_{\rm ion} \sigma_{\rm ion}$. The results yield a cell constant $k = 0.36 \pm 0.01 ~ \rm cm^{-1}$, as shown in the inset of Fig.~\ref{Sfig:calibration}.\\

\newpage

\section{Phase diagram of CMC-CB hydrogels viscoelastic properties }
\begin{figure}[!h]
    \centering
    \includegraphics[width=0.5\linewidth]{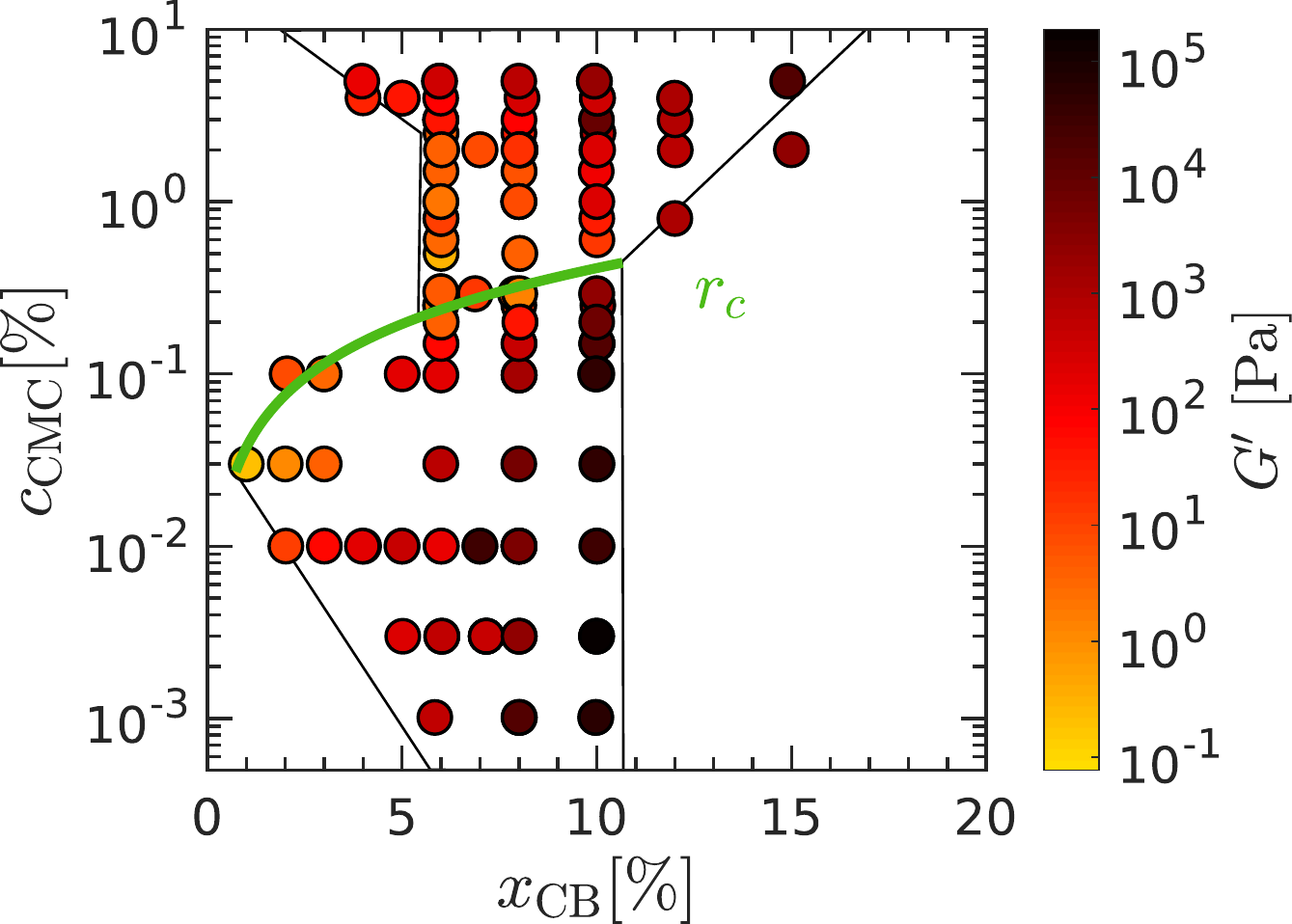}
    \caption{Phase diagram of CMC-CB hydrogels elastic modulus $G'$ measured at $f=1$~Hz and reported as a function of the CB solid weight fraction $x_{\rm CB}$ and the CMC weight fraction $c_{\rm CMC}$. The green curve marks the limit $r=r_c$ that separates two different types of mechanical responses and microstructures.}
    \label{Sfig:diagramme_gseconde}
\end{figure}

\begin{figure}[!h]
    \centering
    \includegraphics[width=0.5\linewidth]{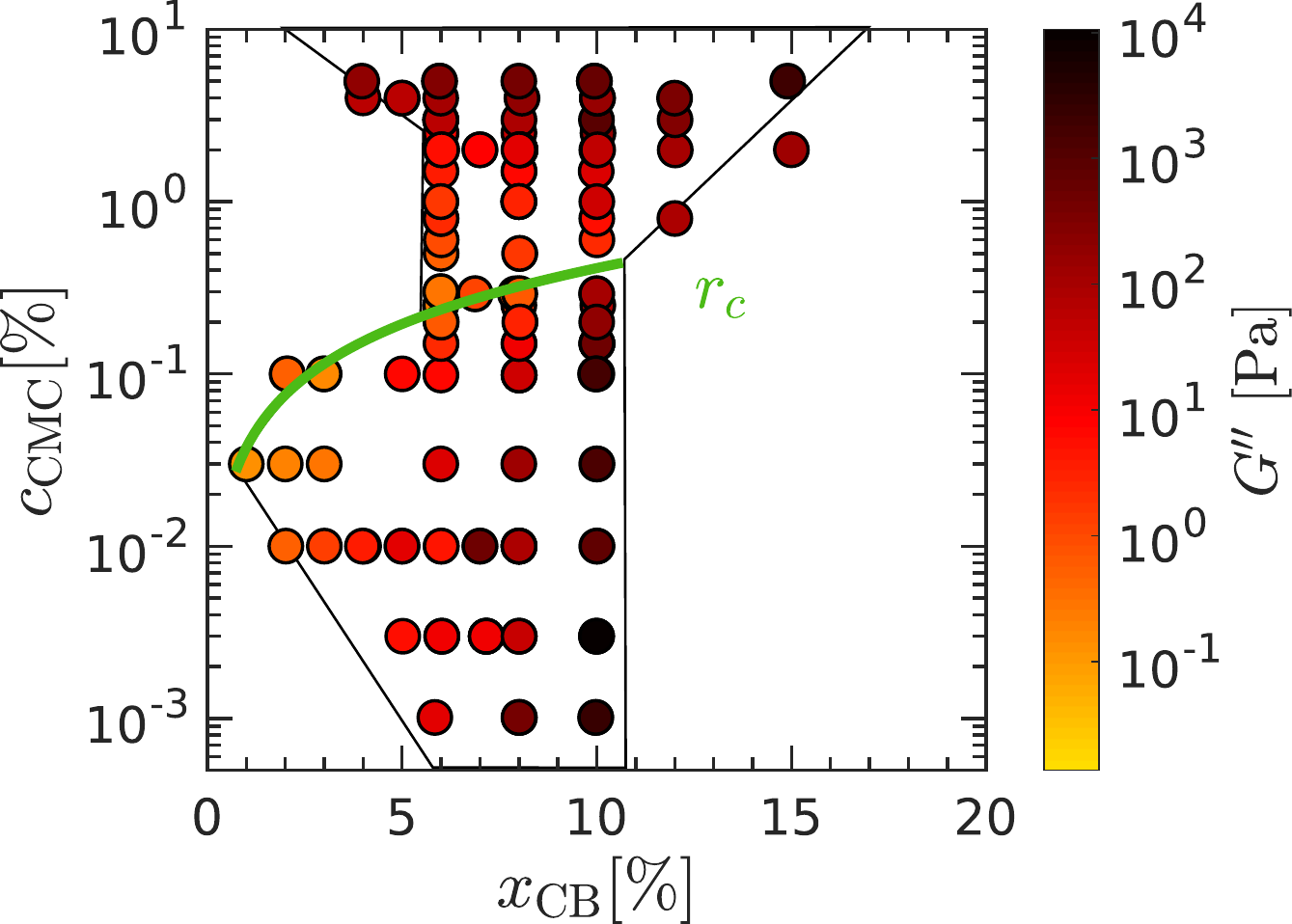}
    \caption{Phase diagram of CMC-CB hydrogels viscous modulus $G''$ measured at $f=1$~Hz and reported as a function of the CB solid weight fraction $x_{\rm CB}$ and the CMC weight fraction $c_{\rm CMC}$. The green curve marks the limit $r=r_c$ that separates two different types of mechanical responses and microstructures.}
    \label{Sfig:diagramme_tandelta}
\end{figure}

Viscoelastic properties measured 1200~s after the end of the rejuvenation step, and reported as 2D maps of the elastic modulus and viscous modulus in the plane ($x_{\rm CB}$, $c_{\rm CMC}$) in Fig.~\ref{Sfig:diagramme_gseconde} and Fig.~\ref{Sfig:diagramme_tandelta}, respectively. The corresponding loss factor $\tan \delta =G''/G'$ is shown in Fig.~3 in the main text.
%These two features exhibit two clear distinctive behavior depending on the type of gel. Namely the loss factor is at least ten times smaller for the gels with $r<r_c$ than for the gels with $r > r_c$.

\clearpage

\section{Justification for the scaling of $G'$ with $c_{\rm CMC}$ for $r < r_c$}
\begin{figure*}[h!]
    \centering
    \includegraphics[width=0.8\linewidth]{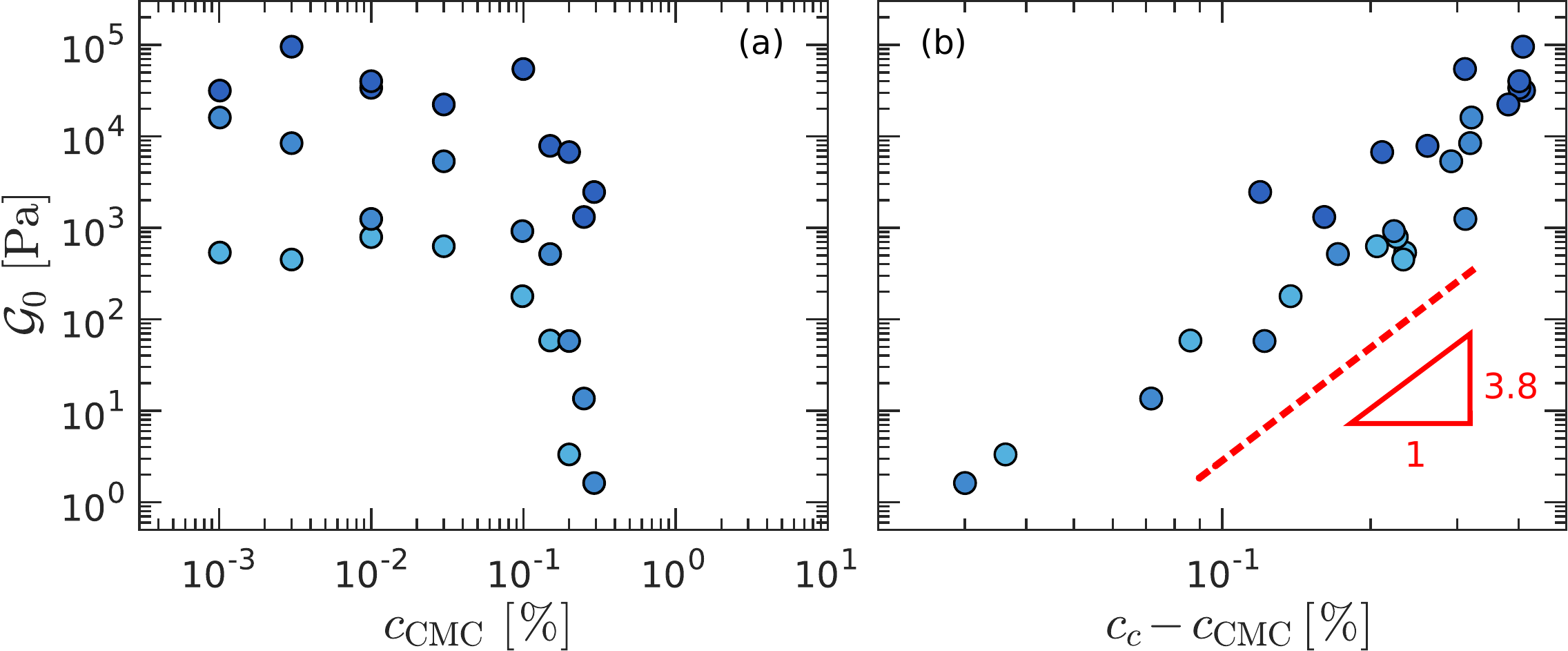}
    \caption{ (a) Elastic modulus $G^\prime$ at $f=1$~Hz as a function of $c_{\rm CMC} $ for different CB contents $x_{\rm CB}=$ 6\% (\textcolor{blue6}{$\bullet$}), 8\% (\textcolor{blue8}{$\bullet$}), and 10\% (\textcolor{blue10}{$\bullet$}). (b) Same data plotted as a function of  $c_c - c_{\rm CMC} $ where the critical CMC concentration $c_c$ was computed using $r_c = 0.037$ leading to $c_c = 0.24, \ 0.32, \ 0.41 \% $ for resp. $x_{\rm CB} = 6, \ 8, \ 10 \%$. The red dotted line indicates a slope of $3.8$, which is expected according to the scaling in Fig.~\ref{fig4}.} 
    \label{Sfig:Gprime_vs_cmc_c}
\end{figure*}

The elastic modulus of CMC-CB hydrogels for $r < r_c$ is a decreasing function of the CMC concentration, as shown in Fig.~\ref{Sfig:Gprime_vs_cmc_c}(a). To highlight the power-law scaling between the elastic modulus and the CMC concentration, the horizontal axis is rescaled using a critical CMC concentration $c_c$. To be consistent with the scaling as a function of $r$ and the unicity of $r_c$, irrespective of the CB content, $c_c$ must depend on $x_{\rm CB}$ following $c_c = r_c x_{\rm CB} / (1 - x_{\rm CB})$. With $r_c = 0.037$ as in Fig.~\ref{Sfig:Gprime_vs_cmc_c}(b), the power-law exponent between $\mathcal{G}_0$ and $c_c - c_{\rm CMC}$ for each CB content are the same and equal to $\xi = 3.8 \pm 0.5$ as expected according to Fig.~\ref{fig4} where $\mathcal{G}_0 \sim (r_c -r)^\xi (x_{\rm CB} - x_c)^\beta = (c_c - c_{\rm CMC})^\xi (1/x_{\rm CB} - 1)^\xi x_{\rm CB}^\beta$. 
One should note that $1/x_{\rm CB} - 1 \simeq 1/x_{\rm CB}$ for $x_{\rm CB} \ll 1 $, such that the elastic modulus scales as $\mathcal{G}_0 \sim (c_c - c_{\rm CMC})^\xi x_{\rm CB}^{\beta - \xi}$.  As shown in Fig.~\ref{fig7}(d), the best power-law fit of $\mathcal{G}_0$ as a function of $x_{\rm CB}$ at fixed CMC concentration $c_{\rm CMC}$ yields an exponent $4.7$, which is compatible with $\beta - \xi = 4.4$.

The fact that the more CB in the solution, the higher $c_c$, emphasizes that the relevant parameter is $r$. Indeed, if the transition between the two regimes were only due to the polymer concentration, increasing the CB weight fraction would increase the effective concentration of polymer everywhere else, leading to a lower $c_c$, which is not what is observed here.\\

\clearpage

\section{Master curve for viscoelastic spectra  for different CMC and CB contents}

\begin{figure*}[h!]
    \centering
    \includegraphics[width=0.95\linewidth]{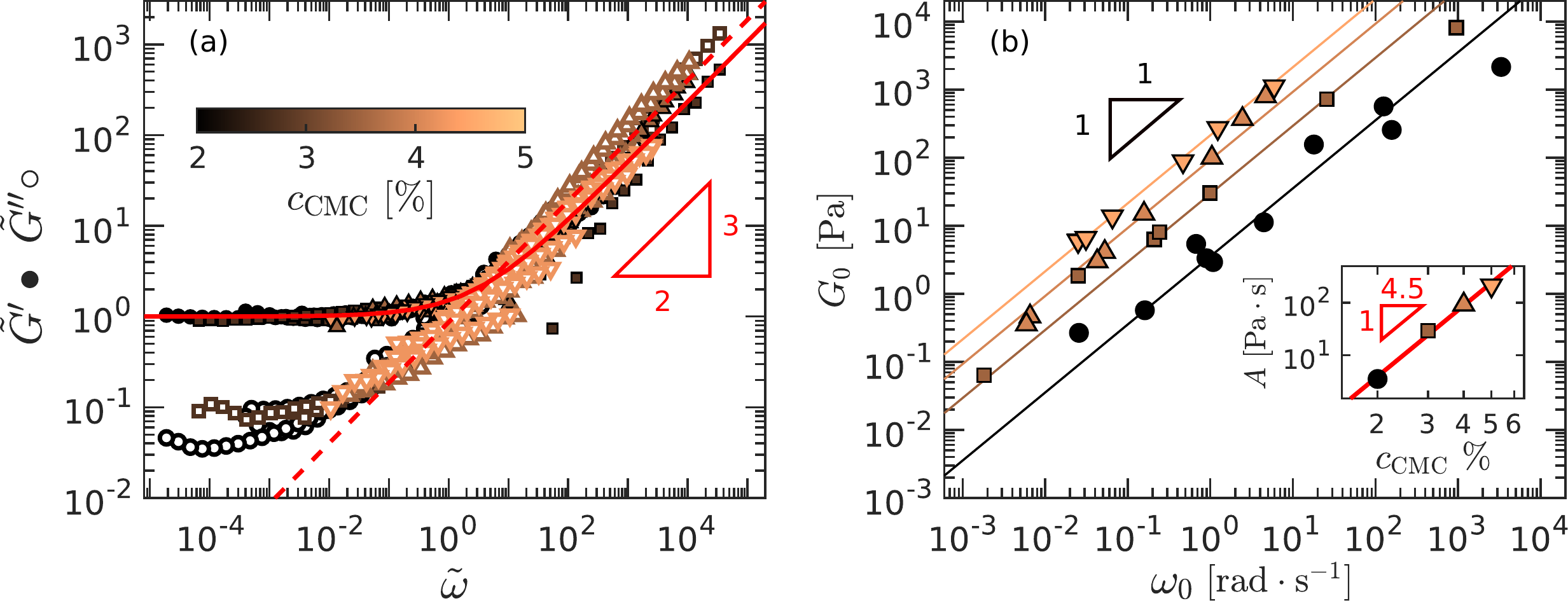}
    \caption{(a) Master curve for the frequency dependence of the viscoelastic moduli of CB-CMC hydrogels for different CMC concentrations [$c_{\rm CMC} = 2~\% \ (\textcolor{cmc2}{\bullet})$, $3~\% \ (\textcolor{cmc3}{\blacksquare})$, $ 4~\%  \ (\textcolor{cmc4}{\blacktriangle})$, {\rm and} $ 5~\% \ (\textcolor{cmc5}{\blacktriangledown})$] and different CB contents ranging between 5~\% and 15~\%. The red curves correspond to the normalized FKV model $\tilde{G}^\star = 1 + \left( i \tilde{\omega} \right)^\alpha$ with $\alpha = 2/3$. (b) Scaling between the two normalization factors $G_0$ and $\omega_0$. The solid lines show the best linear fits of the data $G_0 = A\omega_0$. Inset: $A$ vs. $c_{\rm CMC}$ best described by a power-law fit with an exponent $4.5\pm 0.8$ shown as a red line.} 
    \label{Sfig:master_KVF_CMC}
\end{figure*}

The master curve reported in Fig.~\ref{fig6} corresponds to a rescaling of the viscoelastic spectrum obtained for different CB contents at fixed CMC concentration. Here we show that such a rescaling holds when varying the CMC concentration. Viscoelastic spectra of CB-CMC hydrogels with $c_{\rm CMC}$ varying between 2~\% and 5~\% and $x_{\rm CB}$ varying between 5~\% and 15~\% can be superimposed on the same master curve as that reported in Fig.~\ref{fig6}, using the modulus ($G_0$) and frequency ($\omega_0$) parameters extracted from the fit of each spectra by a fractional Kelvin-Voigt model (FKV). The general master curve is shown in Fig.~\ref{Sfig:master_KVF_CMC}(a). The modulus $G_0$ grows linearly with the frequency $\omega_0$, with $G_0 = A \omega_0$ as shown in Fig.~\ref{Sfig:master_KVF_CMC}(b). The shift factor $A$ increases as a power-law of the CMC concentration with an exponent of about 4.5 [see inset in  Fig.~\ref{Sfig:master_KVF_CMC}(b)]. 
%Note that this exponent is not related to any of the exponents described in the main text since it correlates time and modulus scales, rather than modulus scales and contents of CMC or CB). 

\clearpage

\section{Dependence of the parameter $c$ with the CB content ($r<r_c$)}

\begin{figure*}[h!]
    \centering
    \includegraphics[width=0.5\linewidth]{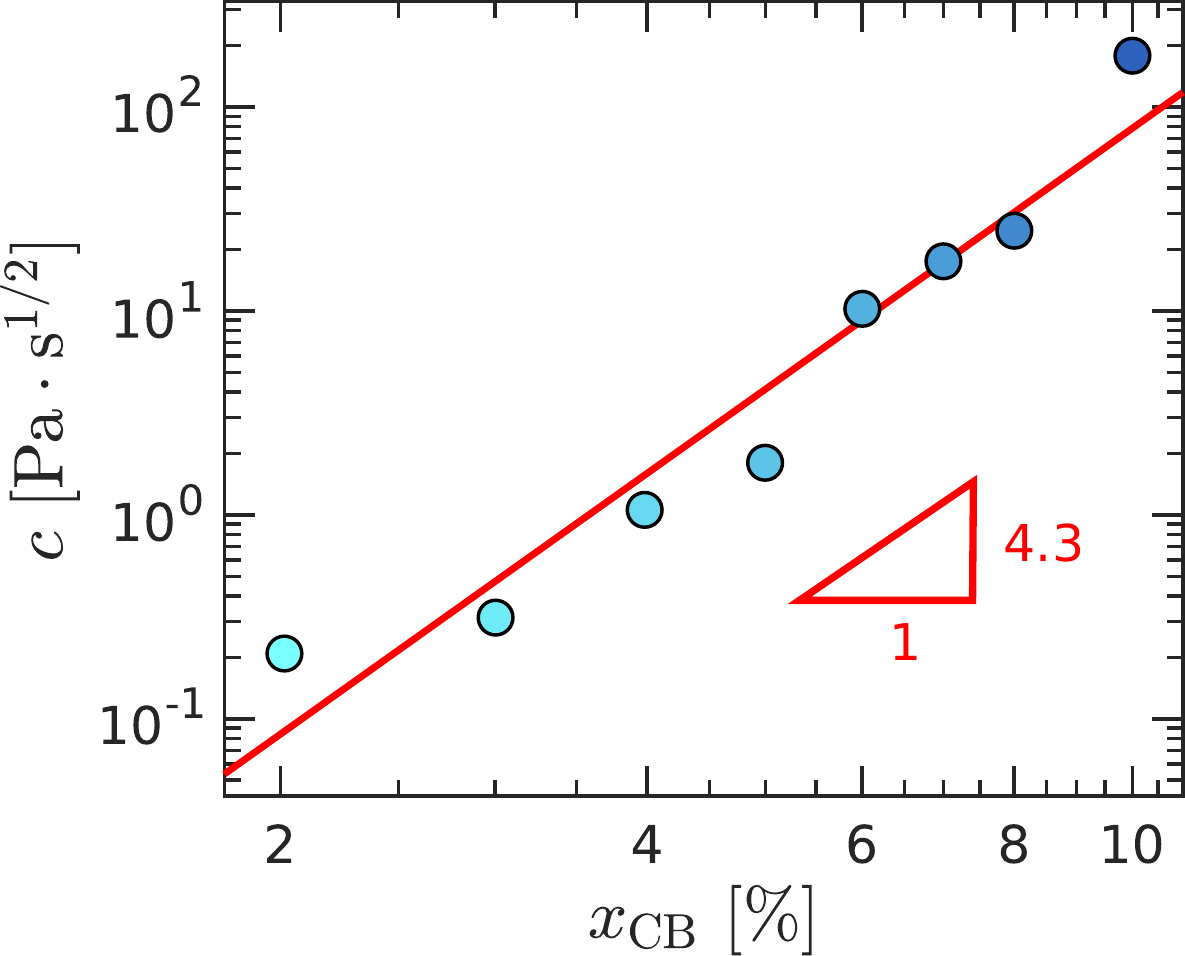}
    \caption{Parameter $c$ as a function of the CB content for CB-CMC hydrogels at fixed CMC concentration $c_{\rm CMC} = 10^{-2}\%$. The red line shows the best power-law fit of the data, which yields an exponent of $4.3\pm0.6$.} 
    \label{Sfig:c_SGR}
\end{figure*}

The modified SGR model\cite{Purnomo:2006} accounts for Brownian motion of the gel microstructure with an additional term $c \omega^{1/2}$ in both the elastic and loss moduli as described by Eqs.~\eqref{eq:SGR1} and~\eqref{eq:SGR2}. For the CB-CMC hydrogels with $r<r_c$, we report in Fig.~\ref{Sfig:c_SGR} a power-law dependence of the parameter $c$ with respect to the CB content. 

\end{document}